\documentclass[twocolumn,superscriptaddress,prb,10pt]{revtex4-1}
\usepackage{verbatim}
\usepackage{braket}
\usepackage{amsmath,amssymb}
\usepackage{tikz}
\usepackage{graphicx}
\usepackage{color}
\usepackage[colorlinks,bookmarks=false,citecolor=blue,linkcolor=red,urlcolor=blue]{hyperref}
\usepackage{times}

\begin{document}

\title{Negativity spectrum of one-dimensional conformal field theories} 

\author{Paola Ruggiero}
\author{Vincenzo Alba}
\author{Pasquale Calabrese}
\affiliation{International School for Advanced Studies (SISSA),
Via Bonomea 265, 34136, Trieste, Italy, 
INFN, Sezione di Trieste}

\date{\today}

\begin{abstract} 

The partial transpose $\rho_A^{T_2}$ of the reduced density matrix $\rho_A$ is the key object to quantify 
the entanglement in mixed states, in particular through the presence of negative eigenvalues in its spectrum. 
Here we derive analytically the distribution of the eigenvalues of $\rho_A^{T_2}$, that we dub {\it negativity spectrum}, 
in the ground sate of gapless one-dimensional systems described by a Conformal Field Theory (CFT), 
focusing  on the case of two adjacent intervals. 
We show that the negativity spectrum is universal and depends only on the central charge of the CFT, 
similarly to the entanglement spectrum. 
The precise form of the negativity spectrum depends on whether the two intervals are in a pure or mixed state, and 
in both cases, a dependence on the sign of the eigenvalues is found. 
This dependence is weak for bulk eigenvalues, whereas it is strong at the spectrum edges. 
We also investigate the scaling of the smallest (negative) and largest (positive) eigenvalues of $\rho_A^{T_2}$. 
We check our results against DMRG simulations for the critical Ising and Heisenberg chains, and against exact results 
for the harmonic chain, finding good agreement for the spectrum, but   
showing that the smallest eigenvalue is affected by very large scaling corrections.

\end{abstract}


\maketitle

\section{Introduction}

\begin{figure}[t]
\includegraphics*[width=0.9\linewidth]{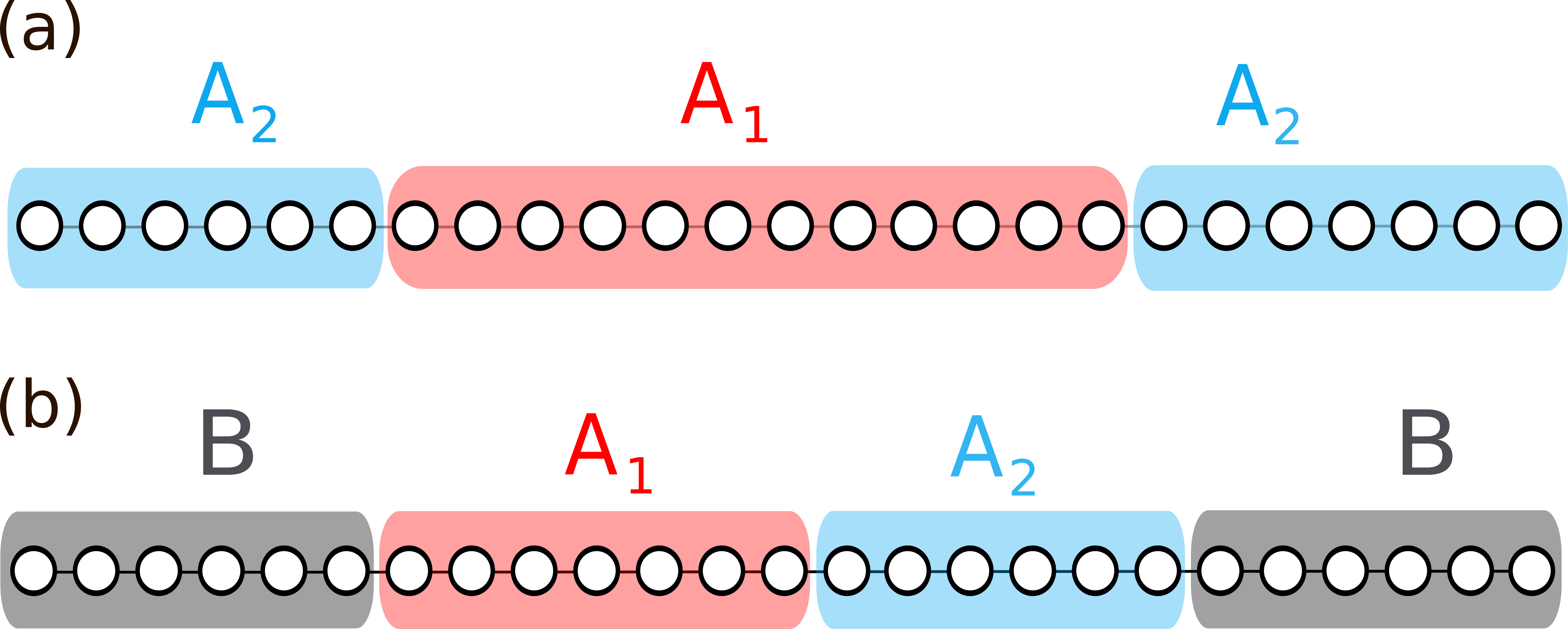}
\caption{Partitions of the 1D pure systems considered in this work. 
Periodic boundary conditions are always implied. 
(a) The bipartition  into two intervals $A_1$ and $A_2=\bar A_1$. 
(b) The tripartition of the chain into two adjacent intervals $A_1$ and $A_2$ with $A\equiv A_1\cup A_2$ 
plus the remainder $B=\bar{A}$. 
In both (a) and (b) the partial transposition is performed with respect to the degrees of freedom in $A_2$. 
}
\label{fig0}
\end{figure}

During recent years devising new tools to detect and characterize the entanglement 
content of quantum many-body systems became a fruitful research theme. For bipartite pure 
states a proper entanglement measure is the entanglement entropy~\cite{amico-2008,
calabrese-2009,eisert-2010,laflorencie-2016}. 
Given a system  in a {\it pure} state $|\psi\rangle$ and a bipartition into two parts $A_1$ and 
$A_2$ (see Figure~\ref{fig0} (a)), the entanglement entropy is defined as 
\begin{equation} 
S_{A_1}\equiv-\textrm{Tr}\rho_{A_1}\ln\rho_{A_1},
\end{equation}
with $\rho_{A_1}\equiv\textrm{Tr}_{A_2}|\psi\rangle\langle\psi|$ being the reduced density 
matrix of $A_1$. For a pure state $S_{A_1}=S_{A_2}$, reflecting that a good measure of 
entanglement is symmetric in $A_1$ and $A_2$. 

However, if a system is in a {\it mixed} state the entanglement entropy is not a good 
entanglement measure, as it is sensitive to both quantum and classical correlations. 
This  happens for finite-temperature systems and if one is 
interested in the mutual entanglement between two non complementary regions of a larger pure system. 
For instance, 
given the tripartition of a system as $A_1\cup A_2\cup B$  (illustrated in Figure~\ref{fig0} 
(b)), with $A\equiv A_1\cup A_2$ the region of interest, $S_{A_1\cup A_2}$ 
is not a measure of the entanglement between $A_1$ and $A_2$. 
In these situations a computable entanglement measure 
is the {\it logarithmic negativity}~\cite{peres-1996,zycz-1998,zycz-1999,lee-2000,vidal-2002,plenio-2005}, 
which is defined as the sum of the absolute values of the eigenvalues of the partially transposed reduced 
density matrix $\rho_A^{T_2}$:
\begin{equation}
\label{neg}
{\cal E}\equiv \ln||\rho_A^{T_2}||_1=\ln\textrm{Tr}|\rho^{T_2}_A|. 
\end{equation}
Here $\rho_A^{T_2}$ is defined as $\langle\varphi_1\varphi_2|\rho^{T_2}_A|\varphi'_1
\varphi'_2\rangle\equiv\langle\varphi_1\varphi_2'|\rho_A|\varphi'_1\varphi_2\rangle$, 
with $\{\varphi_1\}$ and $\{\varphi_2\}$ two bases for $A_1$ and $A_2$, respectively. 
The symbol $||\cdot||_1$ denotes the trace norm. 
Crucially, $\rho_A^{T_2}$ has both positive and negative eigenvalues, in contrast 
with the reduced density matrix, which is positive semidefinite.

The scaling behavior of the negativity has been characterized analytically for the ground states of 
quantum critical models whose low energy physics is captured 
by a one dimensional (1D) Conformal Field Theory (CFT)~\cite{calabrese-2012,cct-neg-long,calabrese-2013}. 
In particular, for disconnected intervals the logarithmic negativity encodes information about the full 
operator content of the CFT~\cite{calabrese-2012}, similar to the entanglement entropy~\cite{2int}. 
Remarkably, the negativity is scale invariant at generic quantum critical points~\cite{hannu-2008,
mrpr-09,hannu-2010,calabrese-2012}. Its scaling behavior is also known for finite temperature 
systems~\cite{calabrese-2015}, in CFTs with large central charge~\cite{kpp-14}, disordered spin 
chains~\cite{ruggiero-2016}, out of equilibrium models~\cite{ctc-14,hoogeveen-2015,eisler-2014,
wen-2015}, some holographic~\cite{rr-15} and  massive quantum field theories~\cite{fournier-2015}, 
topologically ordered phases~\cite{lee-2013,castelnovo-2013}, Kondo-like systems~\cite{bayat-2012,
bayat-2014,abab-16}, and Chern-Simons theories~\cite{wen-2016,wen-2016a}. Surprisingly, no analytical 
results are available yet for free-fermion models, in contrast with free bosonic model, for which 
the negativity can be calculated~\cite{audenaert-2002}, also in $d>1$ dimensions~\cite{eisler-2016,
dct-16}. On the numerical side the negativity can be obtained in DMRG~\cite{white-1992,uli1,uli2} simulations~\cite{hannu-2008,calabrese-2013,ruggiero-2016}. 

Despite this intense theoretical effort,  the properties of the eigenvalues of $\rho^{T_2}_A$ 
have not yet been studied (however in this direction see Ref.~\onlinecite{carteret-2016}). In contrast, the study of the eigenvalues of the reduced density matrix 
(the so-called entanglement spectrum) proved to be an extremely powerful theoretical tool to 
analyze topological phases~\cite{lh-08,topo1,topo2,topo3,topo4}, symmetry-broken phases~\cite{metlitski-2010,alba-2013a,
kolley-2013,frerot-2016}, many-body localized phases~\cite{mbl-es}, and to extract CFT data in models at 
quantum critical points or in gapless phases~\cite{calabrese-2008,lauchli-2013}. For instance, in 
Ref.~\onlinecite{calabrese-2008} it has been shown that for conformally invariant systems the 
entanglement spectrum distribution is described by a universal scaling function that depends only on 
the central charge of the underlying CFT. 
This distribution turned out to be a crucial quantity to understand the scaling (with the auxiliary tensor dimension) 
of matrix product states \cite{chisc}.

In this work we start a systematic study of the spectrum of $\rho_A^{T_2}$, that we dub 
{\it negativity spectrum}, for gapless one dimensional models described by a CFT. 
Specifically, here we investigate,  in the ground state of CFTs, the distribution $P(\lambda)$ of the negativity 
spectrum, which is defined as 
\begin{equation}
\label{P}
P(\lambda)\equiv\sum\limits_{i}\delta(\lambda-\lambda_i), 
\end{equation}
with $\lambda_i$ being the eigenvalues of $\rho_A^{T_2}$. 
Using the same techniques developed Ref.~\onlinecite{calabrese-2008} for the entanglement spectrum, 
we derive analytically $P(\lambda)$ for the case of two adjacent intervals as in Figures~\ref{fig0} (a) and (b). 
We show that the negativity spectrum is sensitive to whether the two intervals are in a pure (Figure~\ref{fig0} (a)) 
or in a mixed state (Figure~\ref{fig0} (b)). 
Moreover, the negativity spectrum distribution is universal and depends only on the central charge 
of the CFT via its largest eigenvalue. 
Its functional form (cf. Eqs.~\eqref{pure} and~\eqref{mixed}) is reminiscent of that of the entanglement spectrum distribution. 
In particular,  $P(\lambda)$ depends on the sign of $\lambda$, but this dependence disappears for the 
asymptotically small (in magnitude) eigenvalues, in both the pure and mixed case. 
Our results imply that the ratio between the total number of positive and negative eigenvalues goes to one in the limit 
of large intervals. 
We also investigate the scaling properties of the support of the negativity spectrum, a subject 
that has  attracted some interest in the quantum information community where it 
has been shown \cite{stv-98,rana-2013}  that the eigenvalues of $\rho_A^{T_2}$ are in $[-1/2,1]$.
Here we focus on the smallest (negative) and the largest (positive) eigenvalues of $\rho_A^{T_2}$ 
(spectrum edges) and we show that for both the pure and mixed case, both the edges exhibit the 
same scaling behavior as a function of the intervals length, which we characterize using  CFT results. 
We show that in the limit of large subsystem the support of the negativity spectrum becomes symmetric, 
i.e. the smallest (negative) eigenvalue is minus the largest (positive) one. 
Interestingly, the negative edge exhibits strong scaling corrections. 
Finally, we provide accurate checks of our results in microscopic models using DMRG simulations, finding always 
excellent agreement. 

The manuscript is organized as follows. In section~\ref{sec::cft} we derive analytically 
$P(\lambda)$ using CFT results, for two adjacent intervals in a pure state in 
subsection~\ref{sec::pure}, and in subsection~\ref{sec::mixed} for the two intervals in 
a mixed state. These are compared with DMRG simulations for the critical transverse field 
Ising chain and the spin-$1/2$ isotropic Heisenberg chain ($XXX$ chain) in 
section~\ref{sec::dmrg}. In subsection~\ref{sec::edges} we discuss the scaling behavior of 
the support of the negativity spectrum. 
We also present exact numerical data for the harmonic chain. 
We conclude in section~\ref{sec::concl}.

\section{Negativity spectrum: CFT results} 
\label{sec::cft}

In this section we derive analytically the distribution of the eigenvalues of the partially 
transposed reduced density matrix (negativity spectrum). 

\subsection{The moment problem}
\label{sec::method}

The negativity spectrum distribution $P(\lambda)$ defined in \eqref{P}
can be reconstructed from the knowledge of its moments $R^{T_2}_n= {\rm Tr} (\rho_A^{T_2})^n$, 
as already done for the entanglement spectrum \cite{calabrese-2008}. 
In terms of $P(\lambda)$, $R_n^{T_2}$ are given by  
\begin{equation}
R^{T_2}_n\equiv\sum_i\lambda_i^n=\int d\lambda\lambda^n P(\lambda), 
\end{equation} 
with $\lambda_i$ being the eigenvalues of $\rho_A^{T_2}$. 

Introducing the Stieltjes transform of $\lambda P(\lambda)$ 
\begin{equation}
\label{fz}
f(z)\equiv\frac{1}{\pi}\sum_{n=1}^\infty R^{T_2}_n z^{-n}=\frac{1}{\pi}\int d\lambda 
\frac{\lambda P(\lambda)}{z-\lambda}, 
\end{equation}
one has \cite{moments}
\begin{equation}
\label{id1}
\lambda P(\lambda)=\lim_{\epsilon\to 0}\textrm{Im} f(\lambda-i\epsilon).
\end{equation}
The distribution $P(\lambda)$ can be effectively reconstructed once the moments $R^{T_2}_n$ are analytically known,   
which is the case for models whose scaling limit is described by a CFT. 
The knowledge of the moments is indeed the starting point to obtain the logarithmic negativity via the 
replica trick~\cite{calabrese-2012} 
\begin{equation}
{\cal E}=\lim_{n_e\to 1}R_{n_e}^{T_2},\quad\textrm{with}\quad n_e\,\,\textrm{even}. 
\end{equation}
It is worth mentioning that, unlike the negativity, the moments $R_n^{T_2}$ can be worked  out analytically in free-fermion 
models~\cite{eisler-2014a,coser-2015,ctc-16,ctc-16b,chang-2016,hw-16} 
and numerically using classical~\cite{alba-2013} and quantum~\cite{chung-2014} Monte Carlo techniques. 
It is also possible in some cases to use numerical extrapolations to obtain the 
negativity from the replica limit of the moments~\cite{dct-15}. 

We recall the reader that this method based on the Stieltjes transform has been used in Ref.~\onlinecite{calabrese-2008} 
to derive the distribution $P_S(\lambda)$ of the entanglement spectrum. The result reads \cite{calabrese-2008}
\begin{multline}
\label{p-intro}
P_S(\lambda)=\delta(\lambda_M-\lambda)\\+\frac{b\theta(\lambda_M-\lambda)}{\lambda 
\sqrt{b\ln(\lambda_M/\lambda)}}I_1(2\sqrt{b\ln(\lambda_M/\lambda)}). 
\end{multline}
where $\lambda_M$ is the largest eigenvalue of $\rho_A$, $b\equiv-\ln\lambda_M$, and 
$I_k(z)$ denotes the modified Bessel functions of the first kind. From~\eqref{p-intro} the mean 
number of eigenvalues $ n_S(\lambda)$ larger than $\lambda$, i.e., the tail 
distribution function, is obtained as~\cite{calabrese-2008} 
\begin{equation}
\label{es-n}
 n_S(\lambda)=\int_{\lambda}^{\lambda_M}d\lambda  P_S(\lambda)=
I_0(2\sqrt{b\ln(\lambda_M/\lambda)}). 
\end{equation}
The effectiveness of this distribution function to describe the entanglement spectrum of gapless 1D models has been 
tested in a few numerical examples \cite{calabrese-2008,pm-10,ahl-12,lr-14,laflorencie-2016}, showing that apart from sizeable finite 
size corrections, Eq. (\ref{es-n}) describes accurately the numerical data for the spectrum.

\subsection{Two intervals in a pure state}
\label{sec::pure}

We start considering the negativity spectrum  for two intervals $A\equiv A_1\cup A_2$ in a pure state  as in Figure~\ref{fig0} (a). 
In this case, the moments of the partial transpose $R_n^{T_2}$ can be written in terms of the moments 
$R_n={\rm Tr}\rho_{A_1}^n$ of the reduced density matrix of $A_1$ as \cite{calabrese-2012,cct-neg-long} 
\begin{equation} 
\label{traces}
R_n^{T_2} =
\begin{cases}
\textrm{Tr} \rho_{A_1}^{n_o}, \qquad \quad n_o \; \text{odd}, \\
(\textrm{Tr} \rho_{A_1}^{n_e/2})^2, \quad n_e \; \text{even}.
\end{cases}
\end{equation}
Importantly, the result depends on the parity of $n$. 
This relation between $R_n$ and $R_n^{T_2}$ signals that in the case of a bipartite pure state the negativity spectrum is 
not independent from the entanglement spectrum. 
Indeed, by using the Schmidt decomposition of an arbitrary bipartite pure state 
it is possible to relate all the eigenvalues of the partially transposed density matrix $\lambda_{i,j}$ 
to the non-zero eigenvalues of the reduced density matrix $\rho_{A_1}$ (or equivalently $\rho_{A_2}$).
It is a simple linear algebra exercise to show the relation~\cite{cct-neg-long,rana-2013,eisler-2014a}

\begin{equation}
\label{pure-eig}
\lambda_{i,j} = \left\{
\begin{array}{cc}
\sqrt{\mu_i \mu_j} &  \quad i<j, \\
\mu_i \qquad  & \quad i = j, \\
-\sqrt{\mu_i \mu_j} &  \quad i>j.
\end{array}
\right.
\end{equation}
The validity of the above relation between $\lambda_{ij}$ and $\mu_j$ can be also inferred from the fact that 
the relations \eqref{traces} force an infinite set of constraints on the eigenvalues: 
since \eqref{pure-eig} satisfy all of these constraints, it must be the only solution of the set of equations \eqref{traces}. 
Notice that the largest (positive) eigenvalue of $\rho_A^{T_2}$ coincides with the largest eigenvalue $\mu_1$ of $\rho_{A_1}$,
while the smallest (negative) eigenvalue of $\rho_A^{T_2}$ is given by $-\sqrt{\mu_1\mu_2}$, where $\mu_{1,2}$ are 
the two largest eigenvalues of $\rho_{A_1}$.

Clearly the relations \eqref{pure-eig} are valid for an arbitrary pure state, but in the case of the ground state of a CFT, 
we can use them to derive the probability distribution $P(\lambda)$ of the $\lambda_{i,j}$ 
from that of $\mu_i$, which, for a CFT, is given by $ P_S(\mu)$ in~\eqref{p-intro}. 
From \eqref{pure-eig}, $P(\lambda)$ can be written as 
\begin{multline}
\label{integ}
P(\lambda)=\sum_{i,j}\delta(\lambda-\lambda_{i,j})=\frac{\textrm{sgn}(\lambda)}{2} P_S(|\lambda|)
\\+\frac{1}{2}\int_0^\infty\!\!\!\!dx \int_0^\infty \!\!\!\!dy \; \delta(|\lambda|-\sqrt{xy}  P_S(x) P_S(y). 
\end{multline}
The $\textrm{sgn}(\lambda)$ function in the first row in~\eqref{integ} is necessary in 
order to correctly take into account the $i=j$ term in~\eqref{pure-eig}. 
Plugging ~\eqref{p-intro} into~\eqref{integ}, the double integral can be explicitly performed (but it is a tedious calculation)
and $P(\lambda)$ can be casted in a form which we will explicitly obtain from the moment problem 
(cf.~\eqref{pure} in the following). 

\subsubsection{Negativity spectrum from the moment problem}

Although \eqref{integ} provides already the final answer for the negativity spectrum for the bipartition of the ground state of a  CFT, 
it is very instructive to recover the same result from the moment problem, especially to set up the calculation for the 
more important and difficult case of two adjacent intervals in a mixed state. 
The key ingredients are the moments of the partial transpose given in \eqref{traces} in terms of the moments of $\rho_{A_1}$.
These, for the ground states of models described by a CFT, in the case of one interval $A_1$ of length $\ell_1$ embedded in 
an infinite system, are given by~\cite{cc-04,cc-rev} 
\begin{equation}
\label{cft-s}
\textrm{Tr}\rho_{A_1}^n=c_n\ell_1^{-\frac{c}{6}(n-\frac{1}{n})}. 
\end{equation}
Here $c$ is the central charge of the CFT and $c_n$ is a non-universal constant. 
Plugging~\eqref{cft-s} in~\eqref{traces}, we rewrite $R^{T_2}_n$ as 
\begin{equation} 
\label{tr-pure}
R_n^{T_2} =
\begin{cases}
c'_{n_o} e^{-b (n_o-\frac{1}{n_o})}, \qquad n_o \; \text{odd}, \\
c'_{n_e} e^{-b(n_e-\frac{4}{n_e})}, \qquad  n_e \; \text{even},
\end{cases}
\end{equation}
where the constants $c'_n$ depend on the parity of $n$ (from~\eqref{traces} and~\eqref{cft-s} one 
has $c_{n_e}'=c^2_{n_e/2}$ and $c'_{n_o}=c_{n_o}$, for $n_e$ even and $n_o$ odd, respectively). 
We have also defined 
\begin{equation} 
b\equiv-\ln \lambda_{M}=\frac{c}6\ln\ell_1+{\rm const}. 
\label{lMp}
\end{equation}
Here $\lambda_M$ is the largest eigenvalue  of $\rho_A^{T_2}$, isolated by 
taking the limit $n\to\infty$ in~\eqref{tr-pure}. 
This limit does not depend on the parity of $n$, making $\lambda_M$ well defined. 
This is true not only for the leading logarithmic term in $\ln\ell_1$, but also for the additive constant, given that 
\begin{equation}
\lim_{n_o\to\infty} \frac{\ln c'_{n_o}}{n_o}= \lim_{n\to\infty} \frac{\ln c_{n}}n
=\lim_{n\to\infty} \frac{\ln c_{n/2}^2}n = \lim_{n_e\to\infty} \frac{\ln c'_{n_e}}{n_e}.
\end{equation} 
For a bipartition of the ground state of a CFT, the largest eigenvalue of $\rho_A^{T_2}$ coincides with the largest eigenvalue 
of $\rho_{A_1}$, in agreement with the general result \eqref{pure-eig}. 

At this point, we have all the ingredients to compute the Stieltjes transform just by plugging~\eqref{tr-pure} into 
the definition of $f(z)$ in~\eqref{fz} and performing the sums over even and odd $n$ separately, obtaining 
\begin{multline}
\label{step}
f(z)=\frac{1}{\pi}\sum_{k=0}^\infty \frac{(4b)^k}{k!}
\sum\limits_{n=1}^\infty\frac{(e^{-b}/z)^{2n}}{(2n)^k} \\+ 
\frac{1}{\pi}\sum_{k=0}^\infty \frac{b^k}{k!}
\sum\limits_{n=1}^\infty\frac{(e^{-b}/z)^{2n-1}}{(2n-1)^k} .
\end{multline}
Here we ignored the presence of the non-universal constants $c'_n$. 
This relies on the assumptions that the $c'_n$ do not change significantly upon varying $n$, as indicated by results in
exactly solvable models~\cite{jin-2004} and numerical works \cite{calabrese-2010}. 
The same assumption has been used in deriving the entanglement spectrum distribution in Ref.~\onlinecite{calabrese-2008}
(and the accuracy of the tail distribution function showed in numerical works \cite{pm-10,ahl-12,lr-14} is a further confirmation
of the plausibility of this assumption).

Remarkably, the two sums in~\eqref{step} can be performed analytically, yielding 
\begin{multline}
\label{step1}
f(z)=\frac{1}{\pi}\sum_{k=0}^\infty \frac{(2b)^k}{k!} \textrm{Li}_k (( e^{-b}/z)^2) \\ 
+\frac{1}{\pi}  \frac{e^{-b}}{z}   \sum_{k=0}^\infty \frac{(b/2)^k}{k!}\Phi (( e^{-b}/z)^2), k, 1/2),
\end{multline}
where $\textrm{Li}_k(y)$ is the polylogarithm function and $\Phi (y, k, a)$ one of its generalization known as Lerch 
transcendent function.
Using the  relation
\begin{equation}
\frac{y}{2^k} \textrm{Im} \Phi  (y^2, k, 1/2) = \frac{\textrm{sgn}(y)}{2} {\rm Im} \,[ \textrm{Li}_k ( |y| )],
\end{equation}
the imaginary part of ~\eqref{step1} reads
\begin{multline}
\label{step1.5}
\textrm{Im} f(z)= \frac{1}{\pi}\sum_{k=0}^\infty \frac{(2b)^k}{k!}  \textrm{Im}\, \textrm{Li}_k (( e^{-b}/z)^2) \\ 
+\frac{1}{\pi} \frac{\textrm{sgn}( e^{-b}/z )}{2}    \sum_{k=0}^\infty \frac{b^k}{k!}   \textrm{Im}\, \textrm{Li}_k ( |e^{-b}/z | ),
\end{multline}
where $\textrm{sgn}(y)\equiv y/|y|$ is the sign function. $\textrm{Li}_k(y)$ is analytic in the complex plane, and it has a branch 
cut on the real axis for $y\ge 1$. Specifically, for $y>1$ and $k\ge 1$ the discontinuity on the cut is $\lim_{\epsilon\to 0}\textrm{Li}_k
(y\pm i\epsilon)=\pm\pi(\ln y)^{k-1}/\Gamma(k)$, with $\Gamma(k)$ the Euler gamma 
function. This implies that 
\begin{multline}
\label{step2}
\lim_{\epsilon\to 0}\textrm{Im} f(\lambda-i\epsilon)=  \lambda_M\delta(
\lambda_{M}-\lambda) + \frac{1}{2\ln(\lambda_{M} /|\lambda|)}\\\times\sum_{k=1}^\infty\frac{[b\ln(
\lambda_M/|\lambda|)]^k}{k!\Gamma(k)}[1+4^k\textrm{sgn}(\lambda)],
\end{multline}
Note that the delta peak $\delta(\lambda_M-\lambda)$ originates from the 
$k=0$ terms in~\eqref{step1.5}. The sum over $k$ in~\eqref{step2} can be performed 
explicitly and from~\eqref{id1}, one obtains $P(\lambda)$ as 
\begin{multline}
\label{pure}
P(\lambda)=\delta(\lambda_{M}-\lambda)\\+
\frac{b\theta(\lambda_{M}-|\lambda|)}{|\lambda|\xi}
\Big[\frac{\textrm{sgn}(\lambda)}{2}I_1(2\xi)+I_1(4\xi)\Big],
\end{multline}
where, again, $I_k(z)$ denotes the modified Bessel function of the first kind, and we 
defined the scaling variable $\xi$ as 
\begin{equation}
\label{xi-def}
\xi\equiv\sqrt{b\ln(\lambda_M/|\lambda|)}. 
\end{equation}
The distribution \eqref{pure} is our final result for the negativity spectrum distribution of a bipartition of the ground state 
of a CFT. 
It is a tedious but elementary exercise to verify that \eqref{pure} coincides with~\eqref{integ}, as it should.

\subsubsection{Some consistency checks}

Before discussing the main  properties and physical consequences of the negativity spectrum distribution \eqref{pure}, it is 
worth to provide  some consistency checks of its correctness. 
A first check of~\eqref{pure} is the normalization condition $\int d\lambda\lambda P(\lambda)=1$. 
Since the term $I_1(4\xi)$ in~\eqref{pure} is odd in the normalisation integral, it gives a vanishing contribution, 
and so we have 
\begin{equation}
\label{norm}
\int d\lambda\lambda P(\lambda)=\lambda_M+\int_{-\lambda_M}^{\lambda_M}d\lambda\frac{b}{2\xi}
I_1(2\xi)=1.
\end{equation}
A less trivial check of~\eqref{pure} is obtained by considering the scaling of the 
logarithmic negativity ${\cal E}= \ln \int d\lambda |\lambda|P(\lambda)$. 
First of all, let us notice that the negativity can be rewritten as 
\begin{equation}
\label{check}
{\cal E}=\ln \lim_{n_e \to 1}\textrm{Tr}\big(\rho_A^{T_2}\big)^{n_e}\simeq \frac{c}2 \ln \ell_1
=-3 \ln \lambda_{M}.
\end{equation}
By parity of the integral, the term $I_1(2\xi)$ in~\eqref{pure} does not contribute to ${\cal E}$. 
Using that $\int_{-\lambda_M}^{\lambda_M} d\lambda I_1(4\xi)b/\xi=\lambda_M^{-3}-\lambda_M$,
one finds that~\eqref{pure} satisfies~\eqref{check}. 

\subsubsection{Properties of the negativity spectrum distribution}

The negativity spectrum distribution \eqref{pure} is reminiscent of the entanglement spectrum \eqref{p-intro}, but it is 
definitively different. 
First of all its support is $[-\lambda_M,\lambda_M]$.
This could have been inferred also in two alternative and easier ways that did not require the knowledge of the 
full negativity spectrum distribution. 
First from the moments $R_n^{T_2}$, the smallest negative eigenvalue $\lambda_m$ can be always obtained from the 
analytic continuations of the even and odd sequences. Indeed, one simply has
\begin{multline}
\label{lmin}
\lim_{n\to\infty}\frac{1}{n}\ln\textrm{Tr}\Big[(\rho_A^{T_2})^{n_e=n}-(\rho_A^{T_2})^{n_o=n}\Big]=\\
= \lim_{n\to\infty}\frac{1}{n}\ln \Big(\sum |\lambda_i|^n -\sum\lambda_i^n\Big)= \\
 \lim_{n\to\infty}\frac{1}{n}\ln \Big(2\sum_{\lambda_i<0} |\lambda_i|^n\Big)
=\ln|\lambda_m|\,,
\end{multline}
where we denoted with $\lambda_i$ the eigenvalues of  $\rho_A^{T_2}$.
Plugging~\eqref{tr-pure} in~\eqref{lmin}, one gets  
\begin{equation}
\label{ddd}
\ln|\lambda_m|=-b=\ln\lambda_M. 
\end{equation}
The second method (which has even a more general validity) simply exploits the relations \eqref{pure-eig}. 
From these we have, as already stated, that the smallest negative eigenvalue is given by 
$\lambda_m=-\sqrt{\mu_1\mu_2}$ where $\mu_{1,2}$ are the two largest eigenvalues of the reduced density matrix. 
We have already seen that generically $\mu_1=\lambda_M$. 
It is also known that for a CFT, the entanglement gap $\mu_1-\mu_2$ closes (i.e. $\mu_1-\mu_2\to 0$) in the 
limit $\ell_1\to\infty$ \cite{calabrese-2008} 
(this result is based on earlier CFT results for the corner transfer matrix spectrum\cite{pt-87}).
From this one concludes $\lambda_m=-\lambda_M$.
However, this second derivation, also shows that the relation $\lambda_m=-\lambda_M$ should be handled with 
a lot of care when comparing with numerics. 
Indeed, it has been shown\cite{calabrese-2008,pt-87} that the entanglement gap closes logarithmically upon increasing 
the interval length $\ell_1$, i.e. $\mu_1-\mu_2\propto 1/\ln \ell_1$. 
Consequently, one has for any finite $\ell_1$ that $|\lambda_m|< \lambda_M$ and 
$\lambda_M-|\lambda_m| \propto 1/\ln \ell_1$ for $\ell_1\to\infty$.
The fact that the gap closes only logarithmically with $\ell_1$ means that in practice one would need extremely large intervals 
in order to see the equality between the largest and the smallest eigenvalues: this is practically impossible to observe in a 
numerical simulation. 

A very important property of the negativity spectrum \eqref{pure} is the presence of a delta peak in $\lambda_M$, 
in complete analogy with the standard entanglement spectrum \eqref{p-intro}.
This means that there exists a single eigenvalue $\lambda_M$ which provides a finite contribution to the negativity 
and to the other quantities obtainable from $P(\lambda)$ (as, e.g., the moments etc.).  
Notice that although the largest and the smallest eigenvalues are equal in the limit of very large interval, their 
contribution to the probability distribution function is very different, since at $\lambda=\lambda_M$ there 
is a delta function, which instead is absent at $\lambda=-\lambda_M$. 
This asymmetry has deep consequences on the various observables such as the number distribution function 
discussed in the following.

Apart from the delta peak, the negativity spectrum \eqref{pure} has  two other terms, one symmetric for $\lambda\to-\lambda$ 
and the other antisymmetric. Notice that $\xi\to 0$ corresponds to $\lambda\to\lambda_M$, while $\xi\to\infty$ to $\lambda\to 0$.
Interestingly, in~\eqref{pure} the only dependence on the sign of $\lambda$ is due to the term $I_1(2\xi)$. 
However, since $I_1(2\xi)/I_1(4\xi)\to 0$ for $\xi\to\infty$, the 
distribution of the small eigenvalues does not depend on their sign. 
This can be understood also from the general relation \eqref{pure-eig} in which
the eigenvalues given by $\pm \sqrt{\mu_i \mu_j}$ are invariant under sign exchange and they are many more than the $\mu_i$'s, in 
the limit of large Hilbert spaces.

\begin{figure}[t]
\includegraphics*[width=0.98\linewidth]{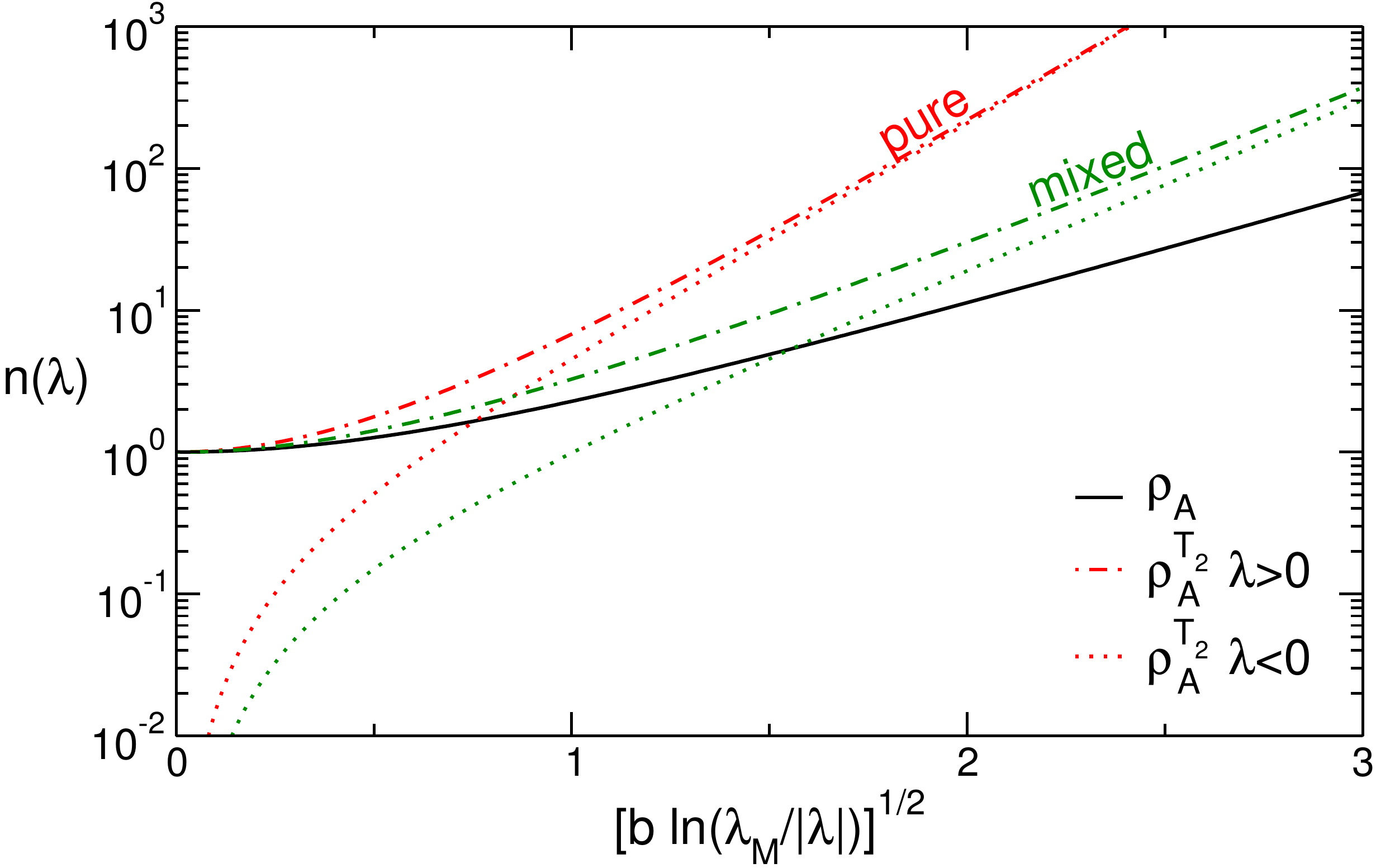}
\caption{ Negativity spectrum of two adjacent intervals: Survey of the CFT 
 results for the number distribution function $n(\lambda)$ plotted versus the 
 scaling variable $\xi\equiv[b\ln(\lambda_M/|\lambda|)]^{1/2}$, with $b\equiv-\ln
 \lambda_M$ and $\lambda_M$ the largest (positive) eigenvalue of $\rho^{T_2}_A$. 
 The continuous line shows for comparison the CFT result for the entanglement 
 spectrum. The dash-dotted and dotted lines denote $n(\lambda)$ for $\lambda>0$ 
 and $\lambda<0$, respectively. The different colors correspond to the case with 
 the two intervals in a pure state and a mixed state. 
}
\label{fig1}
\end{figure}

\subsubsection{The number distribution function}

A crucial observable that is easily obtainable from the negativity spectrum is the mean number of eigenvalues 
larger than a given $\lambda$, i.e. the so-called tail distribution function $n(\lambda)$ which in the following we  refer to 
as {\it number distribution function}. 
The interest in this function comes from the fact that it is a {\it super-universal} smooth function, as already 
known\cite{calabrese-2008} for the entanglement spectrum \eqref{es-n}. 
Indeed, integrating~\eqref{pure}, the number distribution function is given by 
\begin{equation}
\label{pure-n}
n(\lambda)\equiv\int_\lambda^{\lambda_M}d\lambda P(\lambda)=
\frac{1}{2}\big[\textrm{sgn}(\lambda)I_0(2\xi)+I_0(4\xi)\big]. 
\end{equation}
Equation \eqref{pure-n} shows that, also for the negativity spectrum, $n(\lambda)$ is a function of the scaling variable 
$\xi$ only, i.e., there are no free parameters, similarly to the entanglement spectrum \eqref{es-n}. 
In this sense it is super-universal, meaning that it is the same for any CFT. 
Remarkably, the only CFT data appearing in~\eqref{pure} is the central charge, which enters via $\lambda_M$. 
However, when comparing with numerical data, $\lambda_M$ can be fixed from numerics, and consequently 
there is no free/fitting parameter.

There are several interesting properties of the number distribution function worth to be mentioned. 
First of all, the limit for $|\lambda|\to\lambda_M$ (i.e. $\xi\to0$) is very different whether 
one consider positive or negative $\lambda$. 
While for $\lambda>0$, $n(\lambda)\to 1$ as $\lambda\to\lambda_M$, 
for $\lambda<0$, $n(\lambda)\to 0$ as $\lambda\to-\lambda_M$.
This is a straightforward consequence of the presence of the delta peak in $P(\lambda)$ \eqref{pure} 
for $\lambda=\lambda_M$, but not for $\lambda=-\lambda_M$.
In the opposite limit of small absolute value of the eigenvalues,  $n(\lambda)\propto e^{4\xi}/(2\sqrt{2\pi\xi})$ for 
$\xi\to\infty$, independently of the sign of $\lambda$, signaling that the number of small eigenvalues does not depend on 
their sign. 
Finally, $n(\lambda)$ diverges for $\xi\to\infty$  reflecting that in the thermodynamic limit 
the number of eigenvalues of $\rho_A^{T_2}$ is infinite. 
Figure \ref{fig1} reports a plot of the number distribution function versus $\xi$ for both positive and negative $\lambda$.
All the previously listed features should be apparent.

\subsection{Two intervals in a mixed state}
\label{sec::mixed}

We now turn to discuss the case of two adjacent intervals in a mixed state as in Figure~\ref{fig0} (b). 
For two generic intervals $A_1$  and $A_2$, of length $\ell_1$ and $\ell_2$, respectively, 
the scaling of the moments $R_n^{T_2}$ is \cite{calabrese-2012,cct-neg-long} 
\begin{equation}
R^{T_2}_n= c''_n
\begin{cases}
(\ell_1\ell_2)^{-\frac{c}{6}(\frac{n_e}{2}-\frac{2}{n_e})} (\ell_1+\ell_2)^{-\frac{c}{6}(\frac{n_e}{2}+\frac{1}{n_e})}\,,\\
 (\ell_1\ell_2(\ell_1+\ell_2))^{-\frac{c}{12}(n_o-\frac{1}{n_o})} ,
\end{cases}\label{3ptodd}
\end{equation}
where $c$ is again the central charge and the non-universal constants $c''_n$ are analogous to $c'_n$ in (\ref{tr-pure});
they also depend on the parity of $n$, but, as before, are
expected to depend on $n$ in a very weak manner\cite{cct-neg-long} and so will be neglected in the following treatment.
It is convenient to rewrite these moments as 
\begin{equation}
\label{mixed-scal}
R^{T_2}_n\simeq 
\begin{cases}
\ell_1^{-\frac{c}{4}(n_e-\frac{2}{n_e})}\omega^{-\frac{c}{6}(\frac{n_e}{2}-
\frac{2}{n_e})}(1+\omega)^{-\frac{c}{6}(\frac{n_e}{2}+\frac{1}{n_e})},\\
\ell_1^{-\frac{c}{4}(n_o-\frac{1}{n_o})}[\omega(1+\omega)]^{-\frac{c}{12}(n_o-\frac{1}{n_o})},
\end{cases}  
\end{equation}
where $\omega\equiv\ell_2/\ell_1$ is the aspect ratio of the two intervals.
Indeed, from \eqref{mixed-scal} is clear that the largest eigenvalue of $\rho_A^{T_2}$ 
can be extracted by taking the limit $n\to\infty$ which yields the same results from both the even and odd sequences.
This limit leads to 
\begin{multline}
\label{b-new}
b\equiv-\ln\lambda_M=\frac{c}{12}\ln[\ell_1\ell_2(\ell_1+\ell_2)]+{\rm cnst}\\ =
\frac{c}{4}\ln\ell_1+\frac{c}{12}\ln\omega(1+\omega)+{\rm cnst}. 
\end{multline}
We stress that the largest eigenvalue has a different dependence on the central charge compared to the pure case, 
since the prefactor of the logarithm is $c/4$ instead of $c/6$ in \eqref{lMp}.
Notice also that in this case the negativity is not simply a multiple of the logarithm of the largest eigenvalue as in the pure 
case (cf. \eqref{check}), 
but we have (ignoring additive constants)
\begin{multline}
\label{neg-mixed}
{\cal E}=\ln \lim_{n_e \to 1}\textrm{Tr}\big(\rho_A^{T_2}\big)^{n_e}\simeq \frac{c}4 \ln \frac{\ell_1\ell_2}{\ell_1+\ell_2}=\\
= \frac{c}4 \ln \ell_1 +\frac{c}4 \ln \frac{\omega}{\omega+1}=b+\frac{c}6 \ln \frac{\omega}{(\omega+1)^2}.
\end{multline}
At the leading order in $\ell_1$, i.e. ignoring the geometry dependent factor $\omega$, one has ${\cal E}\simeq -\ln \lambda_M$.

The derivation of $P(\lambda)$ from the moments~\eqref{mixed-scal} proceeds as for the pure case 
(see~\ref{sec::pure}) by calculating the 
Stieltjes transform as sum over the moments \eqref{fz}, and taking the limit of its imaginary part as in (\ref{id1}). 
The sums entering in the Stieltjes transform are exactly the same as in the pure case, just with different factors. 
For this reason, we do not report the details of the entire calculation, since after the same algebra as in \ref{sec::pure}, we arrive to 
\begin{multline}
\label{mixed-omega}
P(\lambda)=\delta(\lambda_{M}-\lambda)\\
+\frac{1}{2}\frac{\theta(\lambda_{M}-|\lambda|)}{|\lambda|}\Big[
\frac{b}{\xi}I_1(2\xi)\textrm{sgn}(\lambda)
+\frac{\tilde b}{\tilde\xi}I_1(2\tilde\xi)\Big].
\end{multline}
Here we have introduced the scaling variable $\xi\equiv\sqrt{b\ln(\lambda_M/|\lambda|)}$ (which is the same as in~\eqref{xi-def},
but $\lambda_M$ is different) and also the  auxiliary variable $\tilde\xi$ as 
\begin{equation}
\label{xit-def}
\tilde\xi\equiv\sqrt{\tilde b\ln(\lambda_M/|\lambda|)},\quad\, 
\tilde b\equiv  2b+\frac{c}{6}\ln\frac{\omega}{(1+\omega)^2}. 
\end{equation}
Integrating~\eqref{mixed-omega} between $\lambda$ and $\lambda_M$, we obtain the mean number of eigenvalues $n(\lambda)$ 
larger than $\lambda$
\begin{equation}
\label{n-omega}
n(\lambda)=\frac{1}{2}\Big[I_0(2\xi)\textrm{sgn}(\lambda)+I_0(2\tilde\xi)\Big]. 
\end{equation}
Interestingly, $n(\lambda)$ depends on both $\xi$ and $\tilde\xi$. 
This also implies that the number distribution function is not super-universal as for the pure case 
and for the entanglement spectrum.
However, we will see soon that some major simplifications take place in the limit of large $\ell_1$.

\subsubsection{Some consistency checks}

Before discussing the limit of large $\ell_1$ and the main  properties of the negativity spectrum distribution \eqref{mixed-omega}, 
it is worth to provide some consistency checks. 
As before, a first check of~\eqref{mixed-omega} is the normalization condition $\int d\lambda\lambda P(\lambda)=1$. 
Since the term $I_1(2\tilde\xi)$ in~\eqref{pure} is odd in the normalisation integral, it gives a vanishing contribution.
The remaining integral is identical to \eqref{norm} and provides $\int d\lambda\lambda P(\lambda)=1$.

The second check is given by the scaling of the negativity ${\cal E}=\ln \int d\lambda|\lambda|P(\lambda)$. 
In this case, it is the term $I_1(2\xi)$ to give a vanishing contribution by parity. 
The remaining integral is straightforward and yields 
\begin{equation}
{\cal E}=\ln \int d\lambda|\lambda|P(\lambda)= \tilde b-b= b+\frac{c}{6}\ln\frac{\omega}{(1+\omega)^2},
\end{equation}
which is the expected result in \eqref{neg-mixed}.

The support of the negativity spectrum is $[-\lambda_M,\lambda_M]$ exactly like in the pure case.
The smallest negative eigenvalue $\lambda_m$ can be also obtained by using  \eqref{lmin} 
on the moments \eqref{mixed-scal}. This leads to $\lambda_m=-\lambda_M$  providing another consistency 
check for \eqref{mixed-omega}.

\subsubsection{The limit of large $\ell_1$ and the properties of the negativity spectrum}

We have seen that in general both the probability distribution function $P(\lambda)$ and the resulting number 
distribution $n(\lambda)$ do not depend only on the scaling variable $\xi$, but also on  $\tilde \xi$.
However,  in the limit $\ell_1\to\infty$ many simplifications occur leading to super-universal results.  
First of all, for the largest eigenvalue we have $b=-\ln\lambda_M\to c/4\ln\ell_1$ and also ${\cal E}\to b$,
as clear from ~\eqref{b-new} and \eqref{neg-mixed}. 
Remarkably, this implies that the negativity spectrum distribution does not depend on the geometry of the 
tripartition at the leading order for large lengths of the intervals since from \eqref{xit-def} one has $\tilde \xi\to \sqrt{2} \xi$. 
In this limit, the distribution $P(\lambda)$ simplifies to  
\begin{multline}
\label{mixed}
P(\lambda)=\delta(\lambda_{M}-\lambda)\\+
\frac{b\theta(\lambda_{M}-|\lambda|)}{2|\lambda|\xi}
\Big[\textrm{sgn}(\lambda)I_1(2\xi)+\sqrt{2}I_1(2\sqrt{2}\xi)\Big], 
\end{multline}
whereas $n(\lambda)$ is 
\begin{equation}
\label{mixed-n}
n(\lambda)=\frac{1}{2}\big[\textrm{sgn}(\lambda)I_0(2\xi)+I_0(2\sqrt{2}\xi)\big]. 
\end{equation}
In contrast with~\eqref{n-omega}, in the limit $\ell_1\gg 1$, $n(\lambda)$ is a function of the 
scaling variable $\xi$ only and so it is super universal. 
Note that $P(\lambda)$ for the pure (cf. \eqref{pure}) and mixed case (cf. \eqref{mixed}) have a similar structure, but are 
quantitatively different (the argument of the second Bessel function has a different multiplicative factor). 
Because of this similarity, the most important properties of the negativity spectrum resemble those of the pure case, 
that anyhow we repeat here for completeness. 

As already said, the support of the negativity spectrum is $[-\lambda_M,\lambda_M]$.
However, in analogy with the pure case, the largest and the smallest eigenvalue have a very asymmetrical role, 
because of the the presence of a delta peak in $\lambda_M$, but not at $\lambda_m=-\lambda_M$.
This means that there exists a single eigenvalue $\lambda_M$ which provides a finite contribution to the negativity 
and to the other quantities obtainable from $P(\lambda)$ (as e.g. the moments etc.).  
Oppositely, this is not the case for the smallest eigenvalue.

Moving to the number distribution function \eqref{mixed-n}, the most striking feature is the consequence of the delta 
peak at $\lambda_M$ in $P(\lambda)$. This is indeed the cause of the asymmetry that for 
 $\lambda<0$  one has $n(\lambda)\to 0$ in the limit $|\lambda|\to \lambda_M$, whereas one has 
$n(\lambda)\to 1$, for $\lambda>0$.
Instead the bulk of the negativity spectrum is symmetric.
Indeed, for $\xi\to\infty$ (i.e. for small eigenvalues), one has $n(\lambda)\propto e^{2\sqrt{2}\xi}/(2^{5/4}\sqrt{\pi\xi})$, 
independently from the sign of $\lambda$. 
This is a slower divergence as compared with the pure case (see~\ref{sec::pure}). 
Comparing~\eqref{pure-n} and~\eqref{mixed-n}, one has that in the bulk of the negativity spectrum, i.e., 
for small $|\lambda|$, the scaling relation 
\begin{equation}
n^{\rm pure}\Big(\frac{\xi}{\sqrt{2}}\Big)=n^{\rm mixed}(\xi) ,
\end{equation}
holds.

Figure~\ref{fig1} summarizes all our results for $n(\lambda)$ for both pure and mixed states. 
It reports $n(\lambda)$ versus the scaling variable $\xi\equiv\sqrt{b\ln(\lambda_M/|\lambda|)}$. 
The dash-dotted and dotted lines correspond to $\lambda>0$ and $\lambda<0$, respectively. 
Different colors are used for the case of two intervals in a pure state (cf. \eqref{pure-n}) and in a mixed state (cf. \eqref{mixed-n}). 
The full line shows $n(\lambda)$ for the eigenvalues of the reduced density matrix (entanglement spectrum). 
One has that for any $\lambda$, $n(\lambda)$ is always larger in the pure case. 
Moreover, for $|\lambda|\to 0$ (i.e. $\xi\to\infty$),  $n(\lambda)$ exhibits the same behavior for negative and positive eigenvalues. 
Finally, since the asymptotic behavior of $n(\lambda)$ as $\xi\to\infty$ is independent 
of the sign of $\lambda$, the ratio between the total number of positive 
and negative eigenvalues of $\rho_A^{T_2}$ tends asymptotically to one in both cases.

\subsection{Finite size negativity spectrum}

All the results obtained so far in this section are for finite intervals embedded in infinite one dimensional systems. 
However, one has often to deal with finite systems, especially in numerical simulations. 
Fortunately, all the previous CFT results are straightforwardly generalized to finite systems. 
Indeed, in a CFT, a finite system is obtained by conformally mapping the complex plane to a cylinder.
The net effect of this mapping (for correlations of primary operators and hence for the moments on the reduced density matrix and of 
its partial transpose) is to replace all the lengths with the chord lengths $\ell\to L/\pi\sin(\pi\ell/L)$. 
For the two cases of interest here, this amounts to trivial and unimportant modifications because the probability distribution 
function $P(\lambda)$ and the number distribution function $n(\lambda)$ depend on the lengths only through the maximum 
eigenvalue. Hence, once we replace the maximum eigenvalue of $\rho_A^{T_2}$ with its finite volume counterpart 
equations such as \eqref{pure}, \eqref{pure-n}, \eqref{mixed}, \eqref{mixed-n} still hold. 

In order to be more specific, for the case of a finite periodic system of length $L$ bipartite in two intervals of lengths 
$\ell_1$ and $L-\ell_1$, the maximum eigenvalue of both $\rho_{A_1}$ and $\rho_{A}^{T_2}$ is 
$-\ln \lambda_M= (c/6) \ln [L/\pi \sin (\pi\ell_1/L)]+$ const.
For the case of two adjacent intervals of length $\ell_1$ and $\ell_2$ such that $\ell_1+ \ell_2\neq L$, 
the largest eigenvalue of the $\rho_{A}^{T_2}$ is given by the CFT formula
\begin{multline}
\label{lM-scal-1}
-\ln\lambda_M=\frac{c}{12}\ln\Big[\Big(\frac{L}{\pi}\Big)^3 \sin \big( \frac {\pi \ell_1}{L} \big) \sin \big( \frac {\pi \ell_2}{L} \big) 
\\ \times
\sin \big( \frac {\pi (\ell_1+\ell_2)}{L} \big) 
\Big]+{\rm const}.
\end{multline}
However, in the following, when checking our results for the negativity spectrum with the super-universal CFT forms, 
we will simply fix $\lambda_M$ from the numerical simulations and perform a parameter free comparison with 
 \eqref{pure-n} and \eqref{mixed-n}.

\section{Numerical results}
\label{sec::dmrg}

\begin{figure}[t]
\includegraphics*[width=0.98\linewidth]{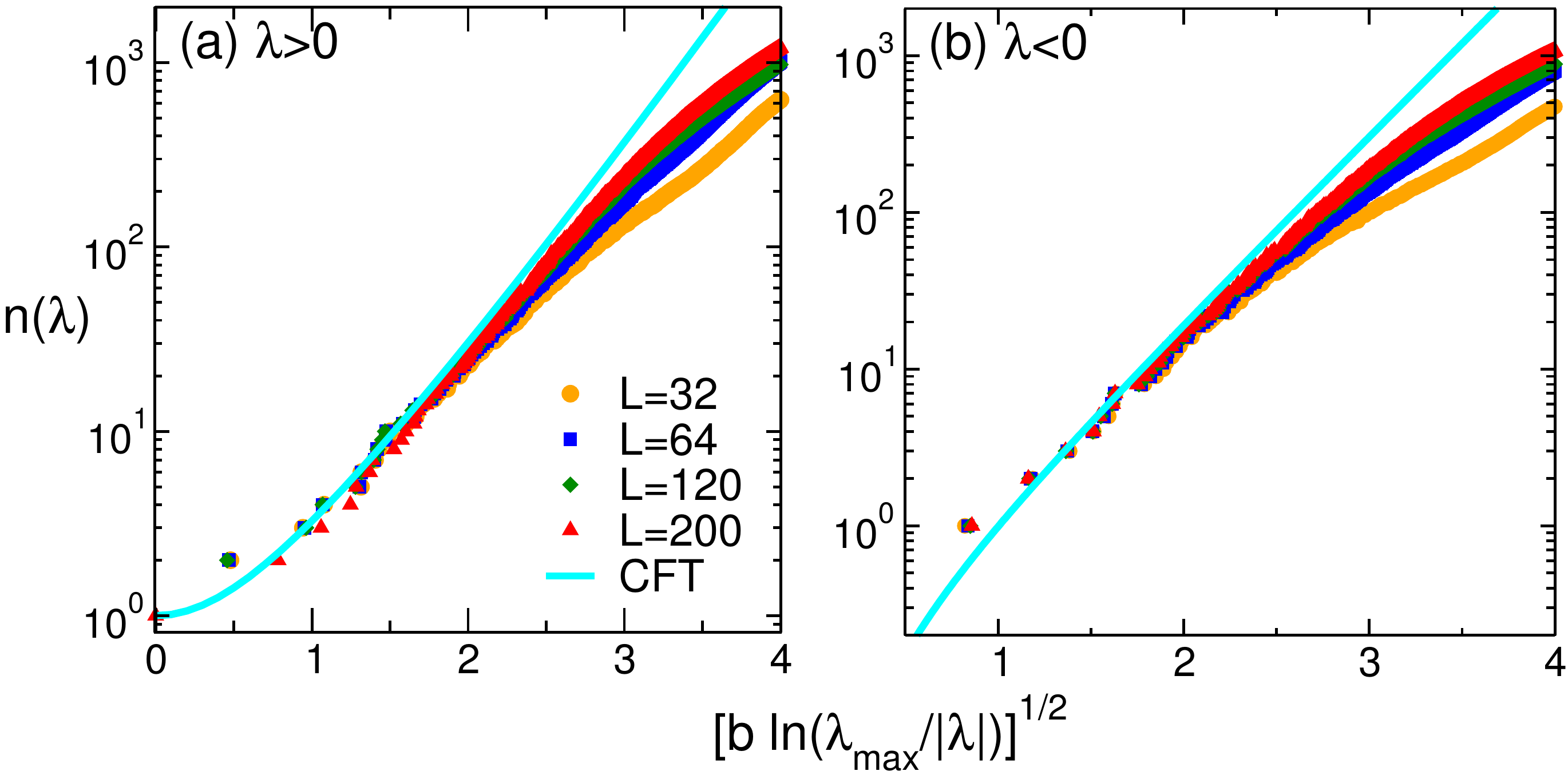}
\caption{Negativity spectrum of two intervals in a pure state.
DMRG results for the critical Ising chain (of sizes $L$) are compared with the CFT prediction 
for the tail distribution $n(\lambda)$ plotted as a function of  
$\xi\equiv[b\ln( \lambda_{M}/|\lambda|)]^{1/2}$ (cf. Eq. \eqref{xi-def}).
The subsystem size is always $\ell=L/2$. 
Panel (a) and (b) report $n(\lambda)$ for $\lambda>0$ and $\lambda<0$, respectively. 
In both panels  the continuous line is the parameter-free CFT prediction. 
}
\label{fig3}
\end{figure}

In this section we provide numerical evidence for the results obtained in 
section~\ref{sec::cft}. We focus on the ground state 
of the critical transverse field Ising chain, and on the spin-$1/2$ isotropic 
Heisenberg chain ($XXX$ chain). We also consider the harmonic chain, for which 
the negativity spectrum can be calculated analytically. 
The Ising chain is defined by the Hamiltonian 
\begin{equation}
\label{ham-is}
{\mathcal H}\equiv -\frac{J}2\sum_{i=1}^L S_i^x S_{i+1}^x-h\sum_{i=1}^LS_i^z. 
\end{equation}
Here $S_i^{x,y,z}\equiv\sigma^{x,y,z}_i$, with $\sigma^{\alpha}_i$ the Pauli matrices, 
are spin-$1/2$ operators acting on site $i$, and $L$ is the chain length. We use periodic 
boundary conditions, identifying sites $1$ and $L+1$ of the chain. We consider the critical 
point at $h=J$, where the low-energy behavior of the model is described by a free 
Majorana fermion, which is a $c=1/2$ conformal field theory. The Heisenberg spin chain 
is instead defined by 
\begin{equation}
\label{ham-xxx}
{\mathcal H}\equiv J\sum_{i=1}^L(S_i^xS_{i+1}^x+S_i^yS_{i+1}^y+S_i^zS_{i+1}^z). 
\end{equation}
The $XXX$ chain is critical, and its low-energy properties are described by the 
compactified free boson (Luttinger liquid), which is a $c=1$ CFT. Again, 
we consider only periodic boundary conditions. 

The Hamiltonian of the periodic harmonic chain is 
\begin{equation}
\label{hc-ham}
{\mathcal H}\equiv\frac{1}{2}\sum_{j=1}^{L}\Big[p_j^2+\Omega q_j^2+(q_{j+1}-q_j)^2\Big]. 
\end{equation}
Here $p_j,q_j$ obey the standard bosonic commutation relations $[q_j,q_k]=[p_j,p_k]=0$, 
$[q_j,p_k]=i\delta_{j,k}$, and $\Omega\in\mathbb{R}$ is a mass parameter. For $\Omega=0$ 
the harmonic chain is critical, and in the scaling limit is described by a $c=1$ free boson. 
Moreover, on the lattice, since~\eqref{hc-ham} is quadratic, 
it can be solved exactly. The partially transposed reduced density matrix has been 
calculated analytically in Ref.~\onlinecite{audenaert-2002} (see also Appendix~\ref{sec::hc}). 
Note that for $\Omega=0$,~\eqref{hc-ham} has a zero mode that leads to divergent expressions. 
For this reason, here we always consider the situation with $\Omega L\ll 1$, choosing 
$\Omega L=10^{-6}$. 

For Ising and Heisenberg spin chains, the partially transposed reduced density matrix $\rho_A^{T_2}$, 
and the negativity spectrum are obtained using DMRG. 
Here we employ the method described in Ref.~\onlinecite{ruggiero-2016}. 
The method relies on the matrix product state (MPS) representation of the ground state 
of~\eqref{ham-is} and~\eqref{ham-xxx}. 
The calculation of the negativity spectrum involves the diagonalization of a $\chi^2\times\chi^2$ matrix, 
with $\chi$ the bond dimension of the MPS. The computational cost is therefore 
$\chi^6$. In our simulations we use $\chi\lesssim 80$, which allows us to simulate system 
sizes up to $L\sim 200$ for the Ising chain, and up to $L\sim 100$ for the $XXX$ chain. 

\subsection{Two intervals in a pure state}
\label{sec::dmrg-pure}

\begin{figure}[t]
\includegraphics*[width=0.98\linewidth]{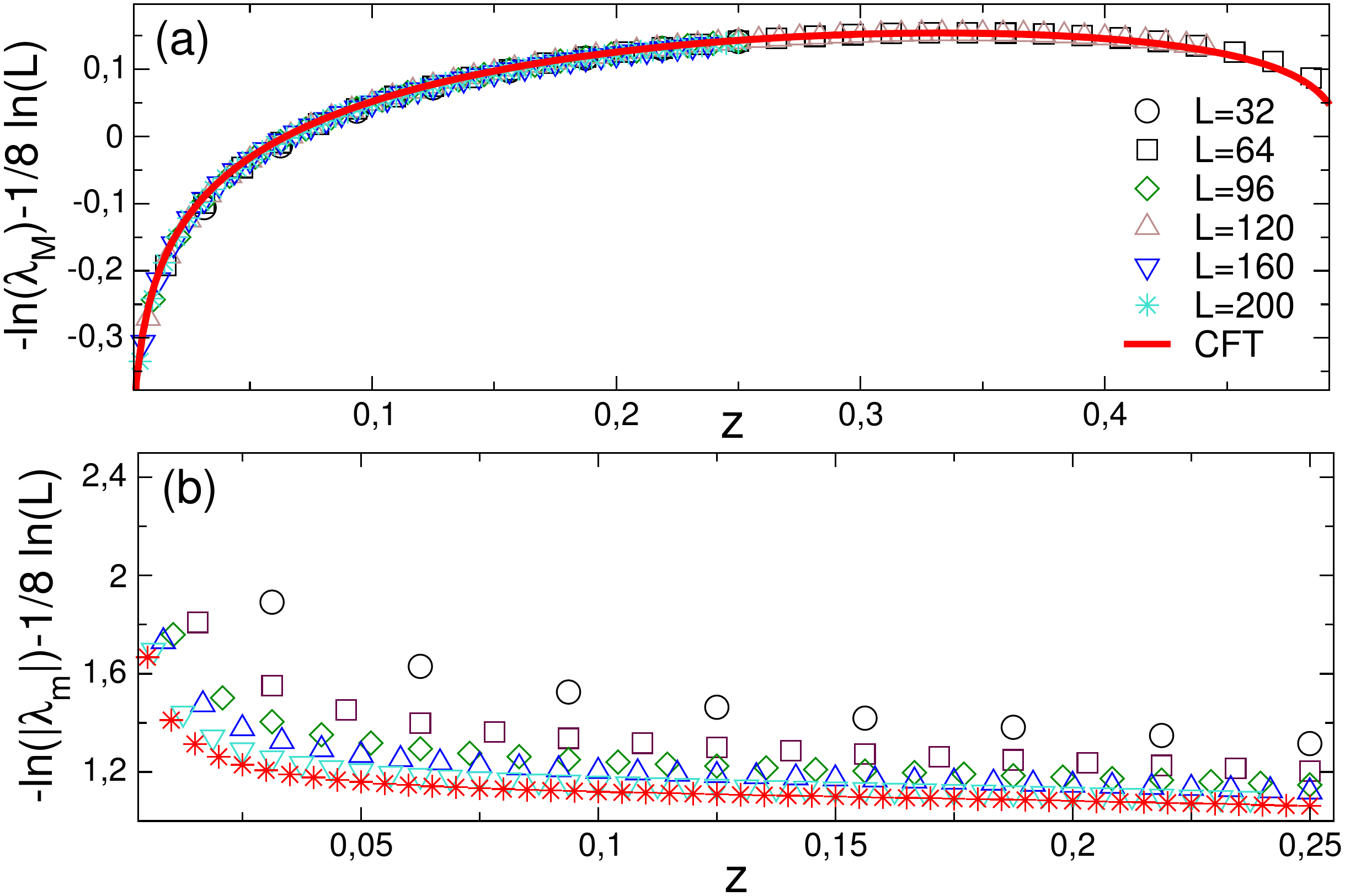}
\caption{Largest and smallest negativity eigenvalues for two adjacent equal intervals of length $\ell$ in a mixed state for a critical 
Ising chain  of length $L$ as function of $z\equiv\ell/L$.
Panel (a): The largest positive eigenvalue $\lambda_M$ of $\rho^{T_2}_A$. 
 The continuous line is the CFT prediction. 
 Panel (b): The smallest (negative) eigenvalue $\lambda_m$ of $\rho^{T_2}_A$.  
}
\label{fig2}
\end{figure}

As we have discussed in Sec. \ref{sec::pure} the negativity spectrum for a system in 
a pure state can be written as a function of the entanglement spectrum of one of the two subsystems. 
Consequently, testing the negativity spectrum of two complementary intervals $A_1$ and $A_2$  (Figure~\ref{fig0} (a))
in the ground state of a CFT is not a new result, but just a further confirmation of the range of validity of the CFT prediction for the 
entanglement spectrum \eqref{es-n}.
It is however instructive to have a look at it, exactly to control the range of validity 
and to test those effects that are not encoded in the CFT predictions such as corrections to the scaling and discreteness 
of the spectrum. 

Here we only provide results for the critical Ising chain considering systems of lengths $L=64,128,256$ 
and a bipartition into two equal intervals of length $\ell_1=\ell_2=L/2$.
We consider the tail distribution $n(\lambda)$ which is plotted 
in Figure~\ref{fig3} versus the scaling variable $\xi\equiv\sqrt{b\ln(\lambda_M/|\lambda|)}$. 
As we already stressed, $n(\lambda)$ depends (via $\xi$) only on $\lambda_M$.
Since we used for $\lambda_M$ the value obtained from the DMRG simulation,
the CFT prediction for $n(\lambda)$ has no free parameters. 
Panels (a) and (b) in  Figure \ref{fig3} are for $\lambda>0$ and $\lambda<0$, respectively. 
The different symbols are the DMRG results for various system sizes, whereas the continuous line is the 
CFT prediction~\eqref{pure-n}. 
The agreement between the CFT prediction and the numerical DMRG data is rather impressive taking into account
that there are no fitting parameters.
There are some small deviations for very small $\xi$ (i.e. for the largest, in absolute value, eigenvalues) which are clearly due to 
the discreteness of the negativity spectrum.  
Then there is a quite large region with $1\lesssim \xi\lesssim 2$ where the agreement is perfect for both 
positive and negative eigenvalues.
For larger $\xi$ (i.e. for very small eigenvalues) sizeable deviations appear. 
These do not come unexpected since they are a consequence of the finiteness of the Hilbert space 
for a block of spin of finite length. Consequently $n(\lambda)$ cannot grow indefinitely as in CFT. 
The same effect is well known and studied already for the entanglement spectrum \cite{calabrese-2008,pm-10,lr-14}.
However, upon increasing $L$ the data exhibit a clear trend toward the CFT prediction, 
confiming that the observed discrepancy is due to scaling corrections and that it should disappear in the limit $\ell\to\infty$.

\subsection{Two intervals in a mixed state: Support of the negativity spectrum}
\label{sec::edges}

We now turn to discuss the negativity of two adjacent intervals $A_1$ and $A_2$ in a 
mixed state as in Figure~\ref{fig0} (b). Before discussing the full negativity spectrum, 
it is instructive to consider the scaling properties of its support. 
In particular, we focus on the scaling behavior of the largest (positive) eigenvalue $\lambda_M$ 
and the smallest (negative) one $\lambda_m$. 
This allows us to control the range of validity of our result and to test those effects that are not encoded
in the CFT prediction, such as scaling corrections.
For simplicity, we restrict ourselves to the situation of two equal-length intervals. 

\begin{figure}[t]
\includegraphics*[width=0.98\linewidth]{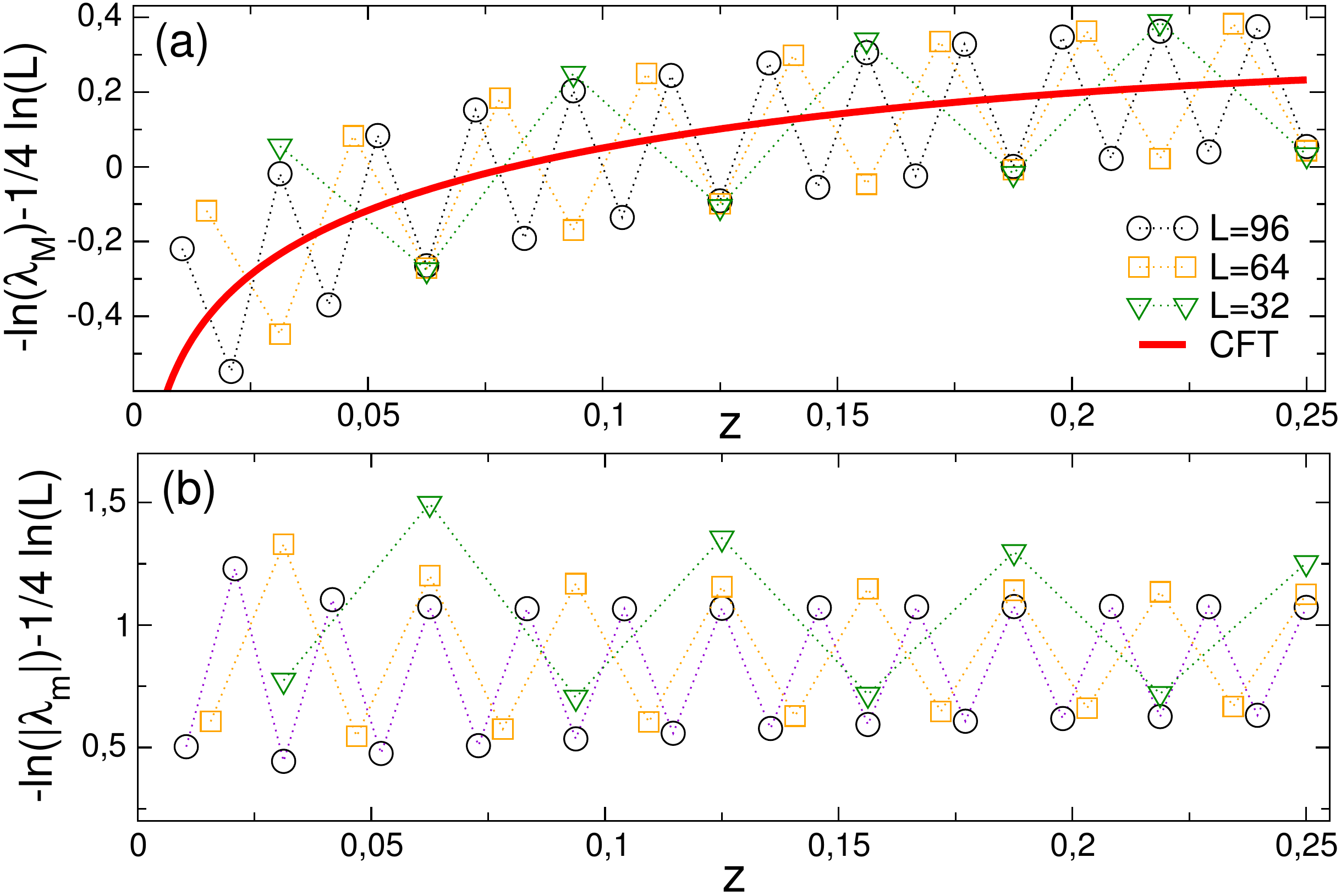}
\caption{ Largest and smallest negativity eigenvalues for two adjacent intervals in a mixed state: 
 Same as in Figure~\ref{fig2} for the $XXX$ chain. 
 Note both in (a) and (b) the presence of oscillating scaling corrections. 
}
\label{fig2a}
\end{figure}

Let us start by discussing the largest eigenvalue. 
An  important consequence of~\eqref{lM-scal-1} (with $\ell_1=\ell_2=\ell$) 
is that the combination $-\ln\lambda_M-c/4\ln L$ is a function of 
$z=\ell/L$ only. For the critical Ising chain this is numerically demonstrated in Figure~\ref{fig2} (a) 
which reports $-\ln\lambda_M-1/8\ln L$ ($c=1/2$ for the Ising chain) versus $0\le z\le 1/2$. 
The different symbols are DMRG data for $L\le 200$. 
The perfect data collapse for all system sizes provides a strong confirmation of~\eqref{lM-scal-1}. 
Moreover, the full line in the Figure is a fit to~\eqref{lM-scal-1}, with the additive constant as the only fitting parameter. 
The agreement with the data is excellent, providing conclusive evidence of the CFT scaling
for the largest eigenvalue of $\rho_A^{T_2}$. 

As we have seen, the smallest negative eigenvalue of $\rho^{T_2}_A$ should scale like the largest positive one $\lambda_M$.
For this reason, to illustrate the scaling behavior of $\lambda_m$,
in Figure~\ref{fig2} (b)  (for the Ising chain and the same tripartition as above)
we report $\ln|\lambda_m|-1/8\ln L$ versus $z=\ell/L$. 
The data do not collapse on a single curve as seen for the largest eigenvalue in panel (a) 
and a quite weak dependence on $z$ is observed. 
This in strikingly different for $\lambda_M$. 
Furthermore, it is clear that the data are not yet asymptotic, suggesting that strong corrections to the scaling 
are present for these values of $L$. 
This does not come unexpected, since we already discussed that strong logarithmic corrections to the scaling 
were expected for $\lambda_m$.

Similar results as in Figure~\ref{fig2} are observed for the $XXX$ chain. 
The CFT scaling~\eqref{lM-scal-1} with $c=1$ is expected to hold for $\lambda_M$. 
Panel (a) in Figure~\ref{fig2a} reports $-\ln\lambda_M-1/4\ln L$ versus $z$. 
In contrast with the Ising case (Figure~\ref{fig2}), strong oscillations with the parity of the intervals length 
$\ell$ are present and should be attributed to the finite-size of the chain. 
Indeed, similar scaling corrections are well known in the literature for the R\'enyi entropies \cite{calabrese-2010,lsca-06,cc-10}, 
and are due to the antiferromagnetic nature of the $XXX$ interaction. 
Moreover, for $-\ln\lambda_M$ of $\rho_A$ these corrections are known to 
decay logarithmically~\cite{calabrese-2010,ce-10} with $\ell$ as $1/\ln\ell$.
Since a similar behavior is expected for the largest eigenvalue of $\rho_A^{T_2}$, this 
explains the very weak dependence on $L$ of the oscillations observed in Figure~\ref{fig2a} (a). 
Still, the CFT result~\eqref{lM-scal-1}, which is shown as full line in Figure~\ref{fig2a} 
(a), captures well the gross behavior of the DMRG data. Finally, in Figure~\ref{fig2a} 
(b) we focus on $\lambda_m$,  reporting $-\ln|\lambda_m|-1/4\ln L$ as a function of $z$, 
for the same chain sizes as in panel (a). 
Interestingly, the same oscillating corrections observed for $\lambda_M$ are present. 
These oscillations prevent to understand the scaling with $L$ of $\lambda_m$
and so the data are even less conclusive than those for the Ising chain in Figure~\ref{fig2}.

\begin{figure}[t]
\includegraphics*[width=0.98\linewidth]{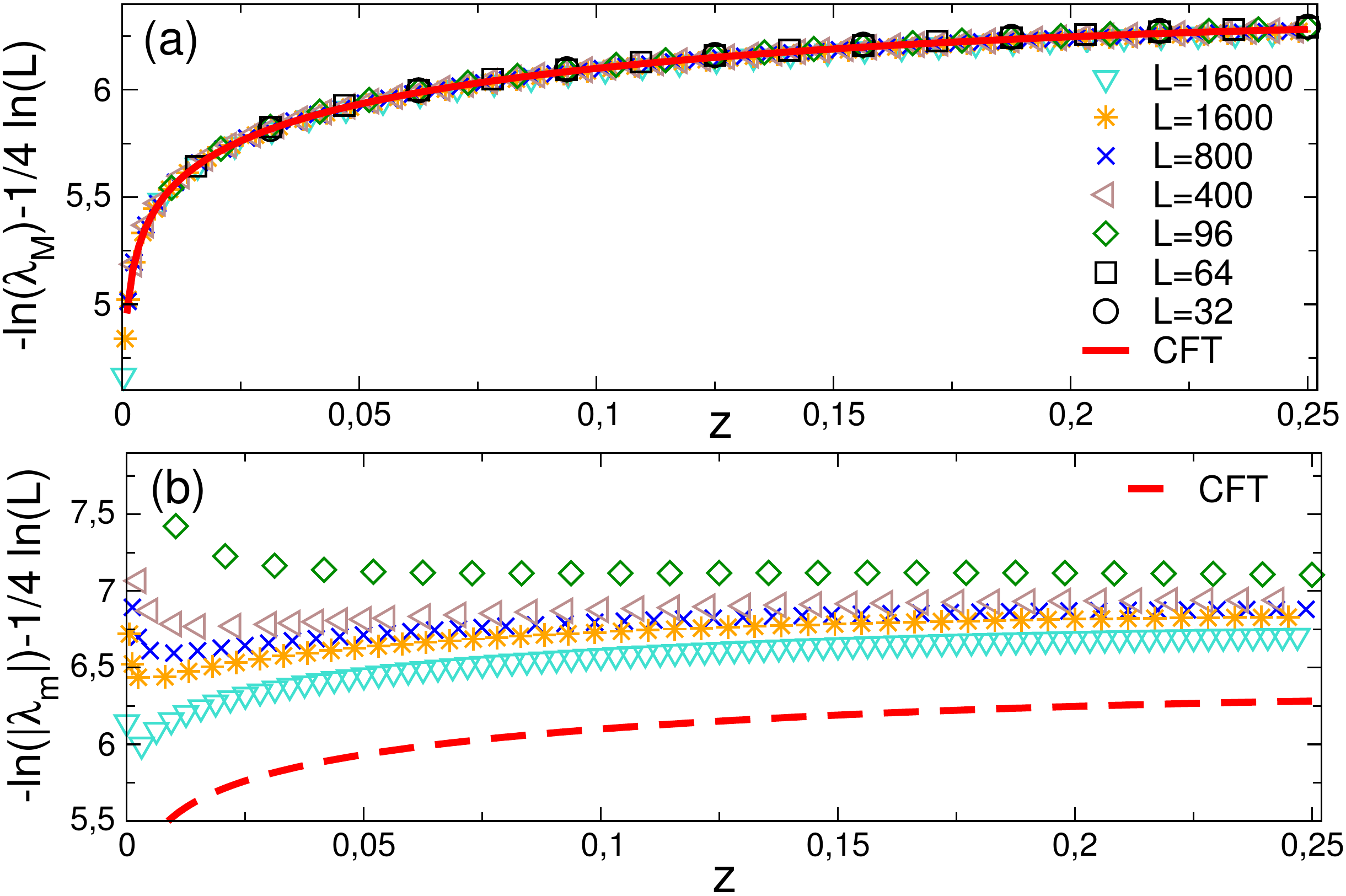}
\caption{Largest and smallest negativity eigenvalues for two adjacent intervals in a mixed state: 
 Same as in Figure~\ref{fig2} for the harmonic chain. 
 Note in panel (b) the very large scaling corrections for $\lambda_m$. 
 The dashed line is the same curve for $-\ln\lambda_M$ as in (a). 
}
\label{fig2b}
\end{figure}

At this point, we do not have yet conclusive data to support the CFT scaling $\lambda_m=-\lambda_M$ for the 
negative edge of the negativity spectrum. We have strong evidence that the data in panels (b) of Figures \ref{fig2} and \ref{fig2a} are 
affected by logarithmic corrections to the scaling. 
Consequently, in order to reveal the true asymptotic behavior, we would need to explore system sizes that are orders of 
magnitude larger than those already considered. This is clearly impossible with DMRG.
For this reason we study the support of the negativity spectrum for the harmonic chain
for which standard techniques for the diagonalization of bosonic quadratic Hamiltonians allowed us to investigate 
chains with $16000$ sites with a minor numerical effort (see the appendix for a review of these techniques).  
The edges of the negativity spectrum for a tripartite harmonic chain (with $\ell_1=\ell_2=\ell$) are reported 
in  Figure~\ref{fig2b}. 
Panel (a) focuses on the largest eigenvalue $\lambda_M$, reporting $\ln\lambda_M-1/4\ln L$ 
versus $z=\ell/L$ for chain up to $L=16000$. 
The agreement between the CFT prediction~\eqref{lM-scal-1} and the data is perfect
(again the only fitting parameter is the additive constant). 
Notice, however, the vertical scale of Figure \ref{fig2b}: we have a very large value, reflecting the fact that for the periodic 
harmonic chain the zero mode produces a large additive constant to the leading logarithmic behavior.  
On the other hand, in panel (b) we plot $\ln|\lambda_m|-1/4\ln L$ versus $z$. 
Strong scaling corrections are still visible at $L=16000$. 
Specifically, while for $L\sim 100$ the data exhibit a ``flat'' behavior as a function of $z$ which is reminiscent of what 
observed in Figure~\ref{fig2} (b) for the Ising chain, 
for larger chains the data become compatible with the CFT scaling~\eqref{lM-scal-1}: it is in fact clear that 
the curve for $L=16000$ is just shifted compared to the asymptotic prediction $|\lambda_m|=\lambda_M$ 
(dashed line in the figure). 
In Appendix~\ref{sec::hc} we report some further evidences that for the harmonic chain in the large $\ell$ limit, 
$|\lambda_m|\to \lambda_M$. 
However, from Figure~\ref{fig2b} (b) it is clear that this can be true 
only for very large chain sizes (comparing the data with the dashed line). 
Once again this fact is fully compatible with the presence of the expected logarithmic corrections to the scaling.

\subsection{The negativity spectrum}
\label{sec::dmrg-ns}

\begin{figure}[t]
\includegraphics*[width=0.98\linewidth]{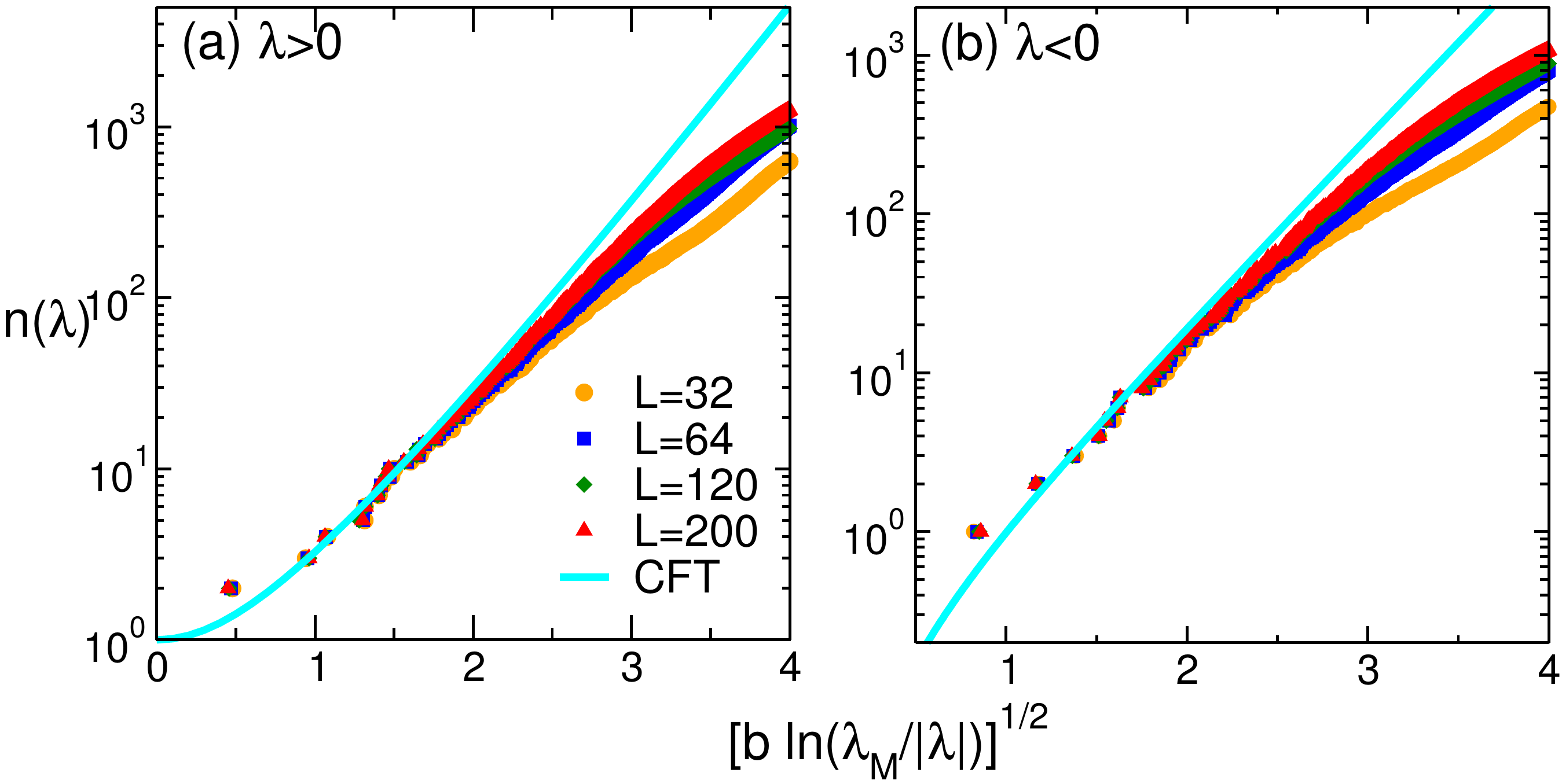}
\caption{ Negativity spectrum of two adjacent equal-length intervals in a mixed state: 
 The number distribution function $n(\lambda)$ plotted as a function of $\xi\equiv[b\ln(
 \lambda_{M}/|\lambda|)]^{1/2}$, with $b\equiv-\ln\lambda_{M}$, and $\lambda_{M}$ 
 the largest positive eigenvalue of $\rho^{T_2}_A$. The symbols are DMRG results for the 
 critical Ising chain for several chain sizes $L$. The subsystem size is always 
 $\ell=L/4$. Panel (a) and (b) plot $n(\lambda)$ for positive and negative values of 
 $\lambda$, respectively. In both panels the continuous line is the CFT prediction. 
}
\label{fig4}
\end{figure}

We finally discuss the negativity spectrum of two adjacent intervals in a mixed 
state as in Figure~\ref{fig0} (b). 
The results for the critical Ising chain are reported in Figure~\ref{fig4}. 
Panels (a) and (b) show the number distribution function $n(\lambda)$ plotted against $\xi\equiv\sqrt{b\ln(\lambda_M/|\lambda|)}$ for 
both $\lambda>0$ and $\lambda<0$. 
The symbols are DMRG data for $L=32-200$. 
We consider two intervals of equal length $\ell=L/4$. 
Similarly to the pure case, in constructing the scaling variable $\xi$ we used for $\lambda_M$ the value from the DMRG simulation, 
so that the CFT prediction for large $\ell$~\eqref{mixed} is super-universal and does not have any free parameter. 
In the two panels the full lines are these super-universal  CFT predictions~\eqref{mixed}. 
The agreement between the DMRG data and the CFT is fairly good. 
As usual in these plots (compare e.g. with Figure \ref{fig3}), some deviations are observed for large $\xi$ (small $|\lambda|$).
As already explained these deviations are due to the finiteness of the Hilbert space of the interval and they are expected to  
disappear in the limit of large $\ell$. Indeed, the observed trend of the data upon increasing $L$ (and hence $\ell=L/4$)
suggests that in the thermodynamic limit the CFT behavior should be recovered.

\begin{figure}[t]
\includegraphics*[width=0.98\linewidth]{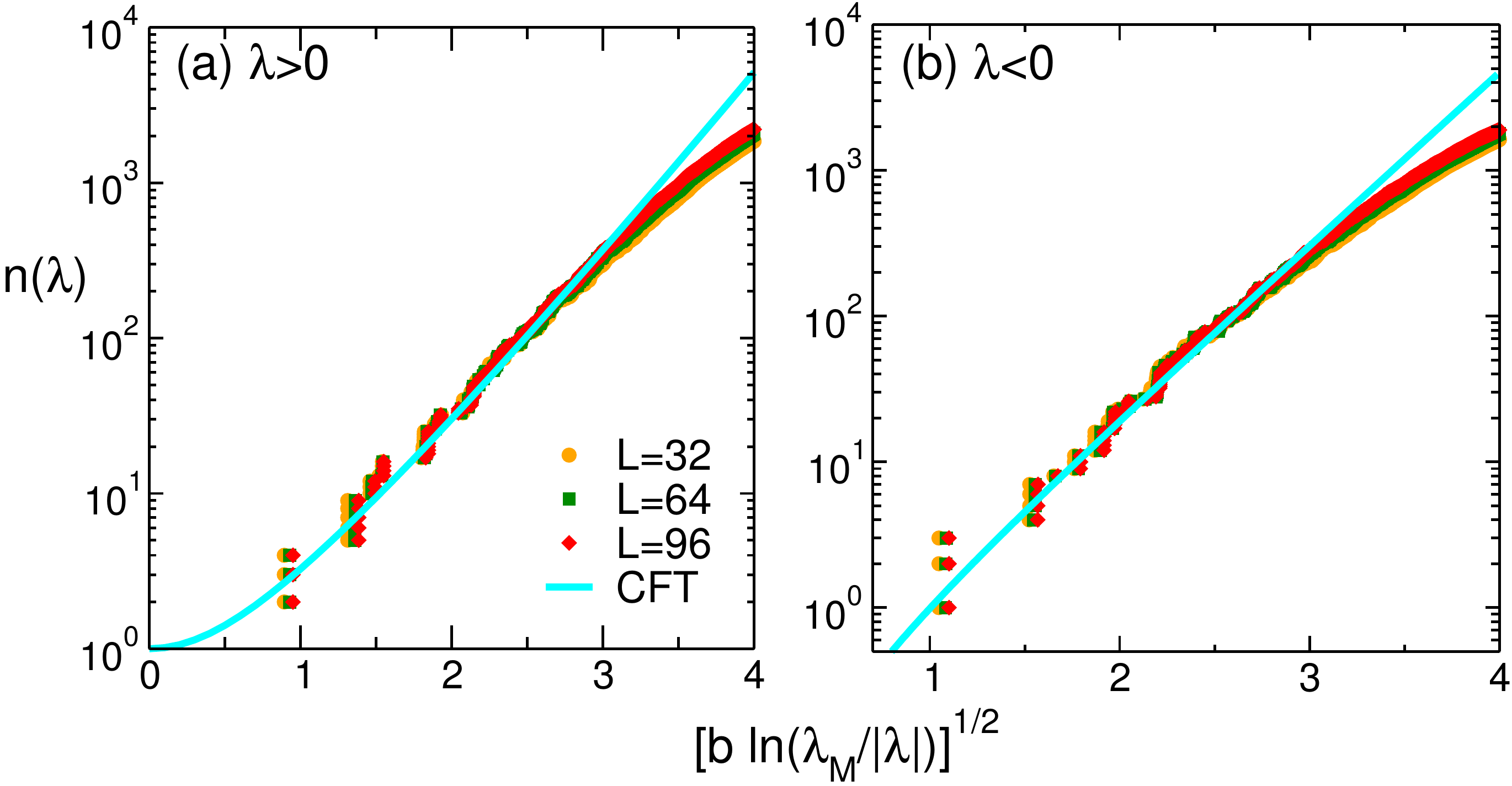}
\caption{ Negativity spectrum of two adjacent intervals in a mixed state: Same as in 
 Figure~\ref{fig4}, for the $XXX$ chain. Note the degeneracy patterns at large  
 $|\lambda|$. 
}
\label{fig5}
\end{figure}

The analogous results for the $XXX$ spin chain for the number distribution function are reported in Figure~\ref{fig5}. 
The symbols are DMRG results now for $L=32-96$. 
The theoretical CFT result is the same as for the Ising chain \eqref{mixed}. 
For both positive and negative values of $\lambda$ the DMRG data are in excellent agreement with the CFT prediction 
(full lines in the Figure). 
It is interesting to observe that at small $\xi$, the negativity spectrum exhibits some intriguing degeneracy structure, which is not 
captured by the CFT result. This is analogous to what observed also for the entanglement 
spectrum in systems with continuous symmetries\cite{calabrese-2008,pm-10}.

Concluding, the results in Figures \ref{fig4} and \ref{fig5} provide a  quite strong evidence for the 
correctness of the CFT negativity spectrum prediction also for the case of two adjacent non-complementary intervals embedded 
in the ground state of model whose low energy features are captured by CFT. 
It is unfortunate that more stringent tests of the CFT prediction cannot be obtained from the study of the harmonic chain. 
Indeed, while for the harmonic chains the methods reported in the appendix allow us to study systems of size $10^4$, 
as we have shown in Figure \ref{fig2b}, the resulting value of $\lambda_M$ is very large because of the presence of the zero mode.
This implies that to get a stringent check of the CFT negativity spectrum, one would require to consider a really huge number of 
eigenvalues of $\rho_A^{T_2}$ which goes beyond our numerical possibilities.

\section{Conclusions \& outlook}
\label{sec::concl}

We investigated both analytically and numerically the distributions of the eigenvalues 
of the partially transposed reduced density matrix (negativity spectrum) in the ground state of one-dimensional gapless systems 
described by a Conformal Field Theory (CFT). Our main results have been already summarized 
in the Introduction. Here we limit to discuss some future research directions originating from our work. 

Clearly, it would be interesting to extend our analysis to the 
case of two {\it disjoint} intervals in a mixed state. It has been demonstrated in 
Ref.~\onlinecite{calabrese-2012} that the negativity of two disjoint intervals decays 
exponentially as a function of their distance. It would be interesting to clarify how this 
behavior is reflected in the negativity spectrum. 
Unfortunately, from the CFT side, this interesting problem is technically prohibitive (if not impossible) because the 
the moments ${\rm Tr} (\rho_A^{T_2})^n$ have a very complicated analytic structure\cite{cct-neg-long}.

Here we focused only on the distribution of the negativity spectrum. It would be enlightening to investigate the fine 
structure of the spectrum, for instance its degeneracy patterns and the eigenvalue spacing. 
This could potentially reveal deeper structures of the underlying CFT, similar to what 
happens for the entanglement spectrum~\cite{calabrese-2008,lauchli-2013}. It would be also 
interesting to study the negative spectrum distribution in gapped phases, as it has been 
done already for the entanglement spectrum in Ref.~\onlinecite{okunishi-1999}. On the 
experimental side, there are recent proposals on how to measure the entanglement 
spectrum in cold-atom experiments~\cite{pichler-2016} (see also Ref.~\onlinecite{carteret-2005}).
It should be possible to extend these ideas to measure the negativity spectrum. Finally, our results could be useful to device 
simpler measures of the entanglement in mixed states. One interesting direction would be to 
focus only on a small portion of the spectrum, for instance exploring region around the 
smallest negative eigenvalue.

\section{Acknowledgments}

We acknowledge very useful discussions with Z.~Zimboras
and V.~Eisler. VA and PC acknowledge support from the ERC under the Starting Grant 279391 EDEQS.

\appendix

\section{Negativity spectrum of the harmonic chain}
\label{sec::hc}

Here we review the calculation of the partially transposed reduced density matrix 
$\rho_A^{T_2}$ for the harmonic chain.
For periodic boundary condition, the Hamiltonian \eqref{hc-ham}  is diagonalized in Fourier space. 
Indeed, defining the operators $\tilde q_k$ and $\tilde p_k$ as 
\begin{align}
\tilde q_k&\equiv\frac{1}{\sqrt{L}}\sum_{s=0}^{L-1} q_se^{-2\pi iks/L},\\
\tilde p_k&\equiv\frac{1}{\sqrt{L}}\sum_{s=0}^{L-1} p_se^{-2\pi iks/L}.
\end{align}
the Hamiltonian~\eqref{hc-ham} becomes 
\begin{equation}
{\mathcal H}=\sum_{k=0}^{L-1}\Omega_k\Big(a^\dagger_ka_k+\frac{1}{2}\Big), 
\end{equation}
where the dispersion $\Omega_k$ is 
\begin{equation}
\label{hc-d}
\Omega_k\equiv\sqrt{\Omega^2+4\sin^2\Big(\frac{\pi k}{L}\Big)}.
\end{equation}
In~\eqref{hc-d} we introduced the creation and annihilation operators $a^\dagger_k$ 
and $a_k$ 
\begin{align}
a_k\equiv&\sqrt{\frac{\Omega_k}{2}}\Big(\tilde q_k+\frac{i}{\Omega_k}\tilde p_k\Big),\\
a^\dagger_k\equiv&\sqrt{\frac{\Omega_k}{2}}\Big(\tilde q_{-k}-\frac{i}{\Omega_k}
\tilde p_{-k}\Big),
\end{align}
satisfying $[a_k,a_{k'}]=[a^\dagger_k,a^\dagger_{k'}]=0$ and $[a_k,a^\dagger_{k'}]=i\delta_{k,k'}$. 
The ground state of the harmonic chain is the vacuum $|0\rangle$ annihilated by $a_k$, i.e., $a_k|0\rangle=0,\forall k$.

To calculate the reduced density matrix $\rho_A$ and its partial transpose $\rho_A^{T_2}$, the correlation matrices 
$\mathbb{Q}_{r,s}\equiv\langle0|q_rq_s|0\rangle$ and $\mathbb{P}_{r,s}\equiv\langle0|p_rp_s|0\rangle$ are required. 
These are obtained by expressing $q_r$ and $p_r$ in terms of $\tilde q_k$ and $\tilde p_k$. The result reads 
\begin{align}
\label{gs-corr}
\mathbb{Q}_{r,s}&=\frac{1}{2L\Omega}+\frac{1}{2L}\sum_{k=1}^{L-1}\frac{1}{\Omega_k}
\cos\Big[\frac{2\pi k(r-s)}{L}\Big]\\
\mathbb{P}_{r,s}&=\frac{1}{2L}\sum_{k=0}^{L-1}\Omega_k\cos\Big[
\frac{2\pi k(r-s)}{L}\Big], 
\end{align}
where in the first row we isolated the divergent term in the limit $\Omega\to 0$, i.e. the zero mode. 

\subsection{Entanglement spectrum}

\begin{figure}[t]
\includegraphics*[width=0.98\linewidth]{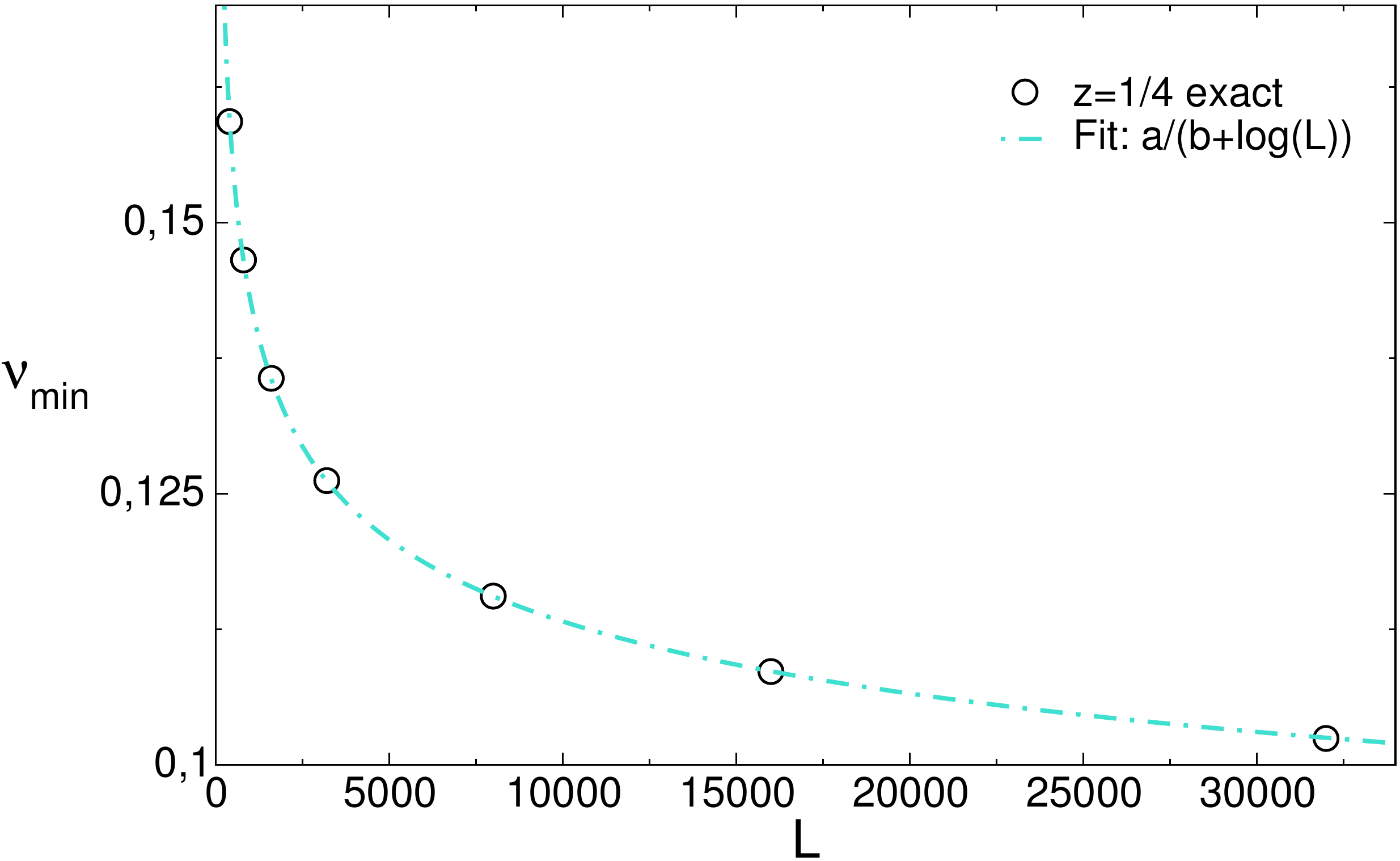}
\caption{  The smallest single particle negativity level $\nu_{min}$ plotted as a function of the chain size $L$. 
The circles are exact results for two adjacent equal-length intervals with 
 $z\equiv\ell/L=1/4$, and $\ell$ being the size of one interval. The dash-dotted 
 line is a fit to $a/(b+\ln(L))$, with $a,b$ the fitting parameters. 
}
\label{fig6}
\end{figure}

The reduced density matrix $\rho_A$ for the ground state of the harmonic chain, and for 
an arbitrary partition of the chain, is fully determined by the correlators $\mathbb{Q}_{r,s}$ 
and $\mathbb{P}_{r,s}$. It can be shown that $\rho_A$ is gaussian and it can be 
written as (see Ref.~\onlinecite{eisler-2009} and references therein) 
\begin{equation}
\label{rho_A-b}
\rho_A=\frac{1}{{\mathcal N}}\exp\Big(-\sum_{j\in A}\epsilon_jb^\dagger_jb_j\Big). 
\end{equation}
Here ${\mathcal N}$ is a normalization factor, and $b_j$ bosonic operators related to the 
original ones $a_j$ in~\eqref{hc-d} by a canonical transformation. In~\eqref{rho_A-b}, 
$\epsilon_j$ are the ``single-particle'' entanglement spectrum levels. Their values is 
fixed by requiring that the expectation values of the correlators of $q_s$ and $p_s$ 
calculated using $\rho_A$ match the corresponding ground state ones (cf.~\eqref{gs-corr}). 
Denoting the restriction of $\mathbb{Q}_{r,s}$ and $\mathbb{P}_{r,s}$ to subsystem $A$ as 
$\mathbb{Q}^A_{r,s}$ and $\mathbb{P}^A_{r,s}$, and given the spectrum $\{\mu_1^2,
\dots,\mu^2_\ell\}$ (with $\ell$ the size of $A$) of $\mathbb{Q}^A\cdot\mathbb{P}^A$, 
one has~\cite{eisler-2009} 
\begin{equation}
\mu_j=\frac{1}{2}\textrm{coth}\frac{\epsilon_j}{2}. 
\end{equation}
Using~\eqref{rho_A-b}, this allows one to write 
\begin{equation}
{\mathcal N}=\prod_{j=1}^\ell\Big(\mu_j+\frac{1}{2}\Big). 
\end{equation}
The spectrum of $\rho_A$ (entanglement spectrum) is obtained by filling in all possible 
ways the single particle levels $\epsilon_j$. Therefore, the entanglement spectrum levels are 
characterize by the occupation numbers $\{\alpha_j\}$ of the bosonic modes. We denote the 
generic level as $\tilde\lambda_{\{\alpha_j\}}$. One has  
\begin{equation}
\tilde\lambda_{\{\alpha_j\}}=\frac{1}{\mathcal N}\prod_{j=1}^\ell\left[\frac{\mu_j-
\frac{1}{2}}{\mu_j+\frac{1}{2}}\right]^{\alpha_j}. 
\end{equation}
Clearly, one obtains $\textrm{Tr}\rho_A^n$ as 
\begin{equation}
\textrm{Tr}\rho_A^n=\prod_{j=1}^\ell\left[\Big(\mu_j+\frac{1}{2}\Big)^n-
\Big(\mu_j-\frac{1}{2}\Big)^n\right]^{-1}.
\end{equation}
The von Neumann entropy is obtained as 
\begin{equation}
S_A=\sum_{j=1}^\ell\Big[\Big(\mu_j+\frac{1}{2}\Big)\ln\Big(\mu_j+\frac{1}{2}
\Big)-\Big(\mu_j-\frac{1}{2}\Big)\ln\Big(\mu_j-\frac{1}{2}\Big)\Big]. 
\end{equation}
%

\subsection{Negativity spectrum}

The partial transpose $\rho^{T_2}_A$ has been constructed in Ref.~\onlinecite{audenaert-2002}. 
The net effect of the partial transposition with respect to part $A_2$ is 
to reverse the momenta corresponding to $A_2$\cite{audenaert-2002} so that the gaussian form of~\eqref{rho_A-b} 
is preserved. The change in the momentum sign is implemented by defining the matrix 
$(\mathbb{P}^A)^{T_2}$ as 
\begin{equation}
(\mathbb{P}^A)^{T_2}\equiv\mathbb{R}^{A_2}\mathbb{P}^A\mathbb{R}^{A_2}, 
\end{equation}
where $[\mathbb{R}^{A_2}]_{r,s}\equiv\delta_{r,s}(-1)^{\delta_{r\in A_2}}$. Basically, 
$\mathbb{R}^{A_2}$ is equal to $\delta_{r,s}$ ($-\delta_{r,s}$) for $r$ in $A_1$ ($A_2$). Note 
that now $\rho_A^{T_2}$ reads 
\begin{equation}
\label{rho_AT2-b}
\rho^{T_2}_A=\frac{1}{{\mathcal N}'}\exp\Big(-\sum_{j\in A}\epsilon_j'b^\dagger_jb_j\Big). 
\end{equation}
The ``single-particle'' negativity spectrum levels $\epsilon_j'$ are obtained from the 
spectrum $\{\nu^2_1,\dots,\nu^2_\ell\}$ of $\mathbb{Q}^A\cdot(\mathbb{P}^A)^{T_2}$ as 
$\nu_j=1/2\textrm{coth}(\epsilon_j'/2)$. Similar to $\rho_A$, one has 
\begin{equation}
{\mathcal N}'=\prod_{j=1}^\ell\Big(\nu_j+\frac{1}{2}\Big).
\end{equation}
The negativity spectrum levels $\lambda_{\{\alpha_j\}}$ are written in terms of the $\nu_j$ as 
\begin{equation}
\label{ns-lev}
\lambda_{\{\alpha_j\}}=\frac{1}{{\mathcal N}'}\prod_{j=1}^\ell\left[\frac{\nu_j-
\frac{1}{2}}{\nu_j+\frac{1}{2}}\right]^{\alpha_j}. 
\end{equation}
One can verify that $0\le\nu_j<\infty$. Clearly, $\nu_j>1/2$ and $\nu_j<1/2$ correspond 
to positive and negative terms in~\eqref{ns-lev}. 
The negativity ${\cal E}$ is obtained from~\eqref{ns-lev} as 
\begin{equation}
{\cal E}=\ln\prod_{j=1}^\ell\max\Big[1,\frac{1}{2\nu_j}\Big]. 
\end{equation}
Clearly, only $\nu_j<1/2$ contribute to ${\cal E}$. 

Finally, since $|(\nu_j-1/2)/(\nu_j+1/2)|<1$ in~\eqref{ns-lev}, the largest positive 
eigenvalue of $\rho_A^{T_2}$ corresponds to all the bosonic modes being unoccupied. On 
the other hand, the smallest negative eigenvalue corresponds to only the mode with the 
smallest $\nu_j$ being occupied. In formulas, one has 
\begin{align}
\label{min}
\lambda_M=&\frac{1}{{\mathcal N}'},\\
\lambda_m=&\lambda_M\frac{\nu_{m}-\frac{1}{2}}{\nu_{m}+\frac{1}{2}},\quad
\textrm{with}\quad\nu_{m}=\min\limits_{j}(\nu_j). 
\end{align}
Notice that $|\lambda_m|<\lambda_M$ for any finite chain. 
The scaling behavior of $\lambda_M$ as a function of the intervals size $\ell$ has 
been presented in Figure~\ref{fig2b} (a). Here we focus on $\nu_{min}$, which determine 
the scaling behavior of $\lambda_m$. Figure~\ref{fig6} reports $\nu_{min}$ as a function 
of the chain size $L$ for $400\le L\le 16000$. The data are for two adjacent equal-length 
intervals with length $\ell=L/4$. 
Clearly, $\nu_{min}$ decreases very slowly upon increasing $L$. To establish the asymptotic equality $\lambda_m\to\lambda_M$, 
it is crucial to understand the asymptotic behavior of $\nu_{min}$ for $L\to\infty$. 
The numerical data in Figure~\ref{fig6} suggests a slow logarithmic decays described by the function  
\begin{equation}
\label{fit}
f(x)=\frac{a}{b+\ln(x)}, 
\end{equation}
which we use to fit the numerical data for $\nu_{\rm min}$.
In~\eqref{fit} the constant $b$ accounts for the dependence of $\ell,L$ on the microscopic cutoff (lattice spacing). 
The behavior~\eqref{fit} has been proved analytically for the single-particle entanglement spectrum 
levels of free fermionic models~\cite{peschel-2004,eisler-2009} 
and conjectured for free bosonic models~\cite{callan-1994,peschel-1999a,peschel-2012,rajabpour-2013} .
Here we only propose it as a fitting function: the result of the fit is shown in Figure \ref{fig6} as dash-dotted line, 
and it is in perfect agreement with the data. 
This suggests that $\nu_{min}\to 0$ in the thermodynamic limit, although larger system sizes would be needed for 
a conclusive evidence. 
From~\eqref{min} this implies that $|\lambda_m|\to\lambda_M$ in the thermodynamic limit, in agreement with the CFT prediction. 
The effect of $\nu_{min}$ is to introduce the logarithmic scaling corrections observed in Figure~\ref{fig2b} (b).


\setcounter{figure}{0}   \renewcommand{\thefigure}{S\arabic{figure}}

\setcounter{equation}{0} \renewcommand{\theequation}{S.\arabic{equation}}


\end{document}